\PassOptionsToPackage{unicode}{hyperref}
\PassOptionsToPackage{hyphens}{url}
\PassOptionsToPackage{dvipsnames,svgnames,x11names}{xcolor}
\documentclass[
  nojss,
  shortnames,
  article]{jss}
\usepackage{xcolor}
\usepackage{amsmath,amssymb}
\usepackage{iftex}
\ifPDFTeX
  \usepackage[T1]{fontenc}
  \usepackage[utf8]{inputenc}
  \usepackage{textcomp} 
\else 
  \usepackage{unicode-math} 
  \defaultfontfeatures{Scale=MatchLowercase}
  \defaultfontfeatures[\rmfamily]{Ligatures=TeX,Scale=1}
\fi
\usepackage{lmodern}
\ifPDFTeX\else
\fi
\IfFileExists{upquote.sty}{\usepackage{upquote}}{}
\IfFileExists{microtype.sty}{
  \usepackage[]{microtype}
  \UseMicrotypeSet[protrusion]{basicmath} 
}{}
\makeatletter
\@ifundefined{KOMAClassName}{
  \IfFileExists{parskip.sty}{%
    \usepackage{parskip}
  }{
    \setlength{\parindent}{0pt}
    \setlength{\parskip}{6pt plus 2pt minus 1pt}}
}{
  \KOMAoptions{parskip=half}}
\makeatother
\makeatletter
\ifx\paragraph\undefined\else
  \let\oldparagraph\paragraph
  \renewcommand{\paragraph}{
    \@ifstar
      \xxxParagraphStar
      \xxxParagraphNoStar
  }
  \newcommand{\xxxParagraphStar}[1]{\oldparagraph*{#1}\mbox{}}
  \newcommand{\xxxParagraphNoStar}[1]{\oldparagraph{#1}\mbox{}}
\fi
\ifx\subparagraph\undefined\else
  \let\oldsubparagraph\subparagraph
  \renewcommand{\subparagraph}{
    \@ifstar
      \xxxSubParagraphStar
      \xxxSubParagraphNoStar
  }
  \newcommand{\xxxSubParagraphStar}[1]{\oldsubparagraph*{#1}\mbox{}}
  \newcommand{\xxxSubParagraphNoStar}[1]{\oldsubparagraph{#1}\mbox{}}
\fi
\makeatother

\usepackage{longtable,booktabs,array}
\usepackage{calc} 
\usepackage{etoolbox}
\makeatletter
\patchcmd\longtable{\par}{\if@noskipsec\mbox{}\fi\par}{}{}
\makeatother
\IfFileExists{footnotehyper.sty}{\usepackage{footnotehyper}}{\usepackage{footnote}}
\makesavenoteenv{longtable}
\usepackage{graphicx}
\makeatletter
\newsavebox\pandoc@box
\newcommand*\pandocbounded[1]{
  \sbox\pandoc@box{#1}%
  \Gscale@div\@tempa{\textheight}{\dimexpr\ht\pandoc@box+\dp\pandoc@box\relax}%
  \Gscale@div\@tempb{\linewidth}{\wd\pandoc@box}%
  \ifdim\@tempb\p@<\@tempa\p@\let\@tempa\@tempb\fi
  \ifdim\@tempa\p@<\p@\scalebox{\@tempa}{\usebox\pandoc@box}%
  \else\usebox{\pandoc@box}%
  \fi%
}
\def\fps@figure{htbp}
\makeatother

\providecommand{\tightlist}{%
  \setlength{\itemsep}{0pt}\setlength{\parskip}{0pt}}

\PassOptionsToPackage{table}{xcolor}

\usepackage{tikz}
\usetikzlibrary{arrows.meta, positioning, calc, fit, backgrounds}
\usepackage{pdflscape}

\usepackage{algpseudocode}

\usepackage{etoolbox}
\AtBeginEnvironment{verbatim}{\small}
\usepackage{orcidlink,thumbpdf,lmodern}

\makeatletter
\@ifpackageloaded{caption}{}{\usepackage{caption}}
\AtBeginDocument{%
\ifdefined\contentsname
  \renewcommand*\contentsname{Table of contents}
\else
  \newcommand\contentsname{Table of contents}
\fi
\ifdefined\listfigurename
  \renewcommand*\listfigurename{List of Figures}
\else
  \newcommand\listfigurename{List of Figures}
\fi
\ifdefined\listtablename
  \renewcommand*\listtablename{List of Tables}
\else
  \newcommand\listtablename{List of Tables}
\fi
\ifdefined\figurename
  \renewcommand*\figurename{Figure}
\else
  \newcommand\figurename{Figure}
\fi
\ifdefined\tablename
  \renewcommand*\tablename{Table}
\else
  \newcommand\tablename{Table}
\fi
}
\@ifpackageloaded{float}{}{\usepackage{float}}
\floatstyle{ruled}
\@ifundefined{c@chapter}{\newfloat{codelisting}{h}{lop}}{\newfloat{codelisting}{h}{lop}[chapter]}
\floatname{codelisting}{Listing}

\makeatother
\makeatletter
\@ifpackageloaded{caption}{}{\usepackage{caption}}
\@ifpackageloaded{subcaption}{}{\usepackage{subcaption}}
\makeatother
\usepackage{bookmark}
\IfFileExists{xurl.sty}{\usepackage{xurl}}{} 
\makeatletter
\@ifundefined{xmpquote}{\newcommand{\xmpquote}[1]{#1}}{}
\makeatother
\hypersetup{
  pdftitle={Fully Homomorphic Encryption for Statistical Modeling},
  pdfauthor={Balasubramanian Narasimhan},
  pdfkeywords={\xmpquote{fully homomorphic
encryption}, \xmpquote{privacy-preserving
computation}, \xmpquote{distributed
statistics}, \xmpquote{R}, \xmpquote{OpenFHE}, \xmpquote{threshold
encryption}},
  colorlinks=true,
  linkcolor={blue},
  filecolor={Maroon},
  citecolor={Blue},
  urlcolor={Blue},
  pdfcreator={LaTeX via pandoc}}

\author{Balasubramanian
Narasimhan~\orcidlink{0000-0001-5852-7639}\\Stanford University}
\Plainauthor{Balasubramanian Narasimhan} 

\title{Fully Homomorphic Encryption for Statistical Modeling}
\Plaintitle{Fully Homomorphic Encryption for Statistical
Modeling} 

\Abstract{Fully homomorphic encryption allows arithmetic to be carried
out on encrypted values, so that a party performing a computation need
not see the data it operates on. This is useful wherever collaborating
parties or sites cannot share data records or computed summaries yet
need to do a joint analysis. We demonstrate the protocols needed to
perform such analyses reproducibly via two packages in \proglang{R}, a
platform widely used for applied statistics. The first, \pkg{openfhe.R},
is an interface to the OpenFHE C++ library, which exposes exact integer
arithmetic (BFV, BGV), approximate real-valued arithmetic (CKKS),
Boolean circuits, and multiparty key generation. The second is
\pkg{homomorpheR}, which builds a small set of master/worker primitives
for multi-party protocols on top of the first. Together, the two let an
ordinary \proglang{R} statistical routine compose with an
encrypted-arithmetic layer at the function-value boundary, provided the
quantity crossing that boundary decomposes as a sum over sites. Two
validation studies are presented: distributed stratified Cox regression,
where \texttt{stats4::mle()} converges through an encrypted channel, and
federated Cox-lasso via convex optimization using consensus ADMM, where
the cross-site update is the only encrypted step. We also discuss
federated similarity retrieval and two-party prediction. Throughout, the
parties are assumed honest-but-curious, and threshold key generation
removes the assumption that any single party holds a usable secret key.}

\Plainkeywords{fully homomorphic encryption, privacy-preserving
computation, distributed statistics, R, OpenFHE, threshold encryption}

\Address{
Balasubramanian Narasimhan\\
Department of Biomedical Data Science, and Department of Statistics\\
Stanford USA\\
E-mail: \email{naras@stanford.edu}\\
\\~

}

\begin{document}
\maketitle

\section{Introduction}\label{sec-intro}

A recurring problem in biomedical research is fitting statistical models
across data held by multiple parties/sites when they cannot share their
patient-level records. For many models, the likelihood, score, and
information decompose as sums over sites, so the fitting algorithm needs
only site-level summaries at each iteration, never the raw rows.
Packages such as \pkg{distcomp}
\citep{NarasimhanRubinGrossBenderskyLavori2017, distcomp} and platforms
such as \pkg{DataSHIELD}
\citep{WolfsonDataSHIELD2010, GayeDataSHIELD2014} implement this
decomposition in \proglang{R}, but the site summaries travel in
cleartext even if the transport mechanism is encrypted: the aggregating
party sees the per-site log-likelihoods, scores, or regression
coefficients, and the privacy story rests on operational and contractual
controls rather than on cryptographic protection.

Fully homomorphic encryption (FHE) offers a way to close this gap. FHE
is a cryptographic technique that allows a third party to perform
arithmetic on encrypted data without seeing the data and without holding
a secret key, obtaining a result that decrypts to what the same
arithmetic on the unencrypted values would give --- exactly, under the
integer schemes, and to a quantifiable approximation under the
real-valued one. Someone must still hold key material in order to
decrypt; the point is that it need not be the party doing the computing,
and under threshold keys it need not be any single party. After
\citet{Gentry2009} gave the first plausible construction, a sequence of
refinements
\citep{BrakerskiVaikuntanathan2014, FanVercauteren2012, CheonKimKimSong2017, ChillottiGamaGeorgievaIzabachene2020}
produced practical schemes for exact integer arithmetic, approximate
real-number arithmetic, and Boolean logic. The OpenFHE C++ library
\citep{AlBadawiOpenFHE2022} gathers modern implementations of these
schemes along with multiparty extensions
\citep{MouchetTroncosoPastorizaBossuatHubaux2021} that distribute the
secret key across several parties so that no single party can decrypt
unilaterally.

We describe two \proglang{R} packages that together make this machinery
available to statisticians:

\begin{itemize}
\tightlist
\item
  \pkg{openfhe.R} --- an \proglang{R} interface to the OpenFHE C++
  library. The package wraps OpenFHE's public-key primitives so users
  can encrypt numeric vectors, operate on the encrypted values with
  ordinary \proglang{R} arithmetic operators (\texttt{+}, \texttt{-},
  \texttt{*}), and decrypt the results.
\item
  \pkg{homomorpheR} --- a higher-level package that builds on
  \pkg{openfhe.R} and provides the master/worker primitives for
  multi-party protocols: the \texttt{Site} and \texttt{Master} classes,
  threshold key generation, and the \texttt{master\_aggregate()} runner.
  It ships worked-example vignettes covering distributed maximum
  likelihood, stratified Cox regression, federated convex optimization,
  privacy-preserving inference, and aggregation; the consensus-ADMM
  driver of Section~\ref{sec-cvxr} is one of those worked protocols,
  built from the package's primitives rather than exported by it.
\end{itemize}

A central observation is that an \proglang{R} routine which consumes
numeric values can compose with the encrypted-arithmetic layer
\emph{without being modified}, whenever the quantity it asks for
decomposes as a sum over sites. The composition occurs at the
function-value boundary: the routine receives a single numeric value (or
a vector of values) and does not need to know how that value was
computed.

Two constraints apply, and both matter in practice. The per-round
computation must be one the scheme supports within its depth and
precision budget --- sums and products of bounded degree, not arbitrary
control flow. And the \emph{analysis} usually has to be arranged so that
such a boundary exists: below, one example introduces a callback,
another decomposes a likelihood additively, and a third is reformulated
as consensus ADMM. What stays unchanged is the statistical solver; the
protocol around it is the work. We focus on two substantial patterns of
composition:

\begin{enumerate}
\def\labelenumi{\arabic{enumi}.}
\tightlist
\item
  \textbf{Function-call optimizers.} \texttt{stats4::mle()} with BFGS,
  or any \texttt{optim()}-style routine. The optimizer asks for the
  value of an objective at a point in \(\mathbb{R}^p\); the callback
  obtains that value through an encrypted master/worker round. In the
  Cox example, \texttt{survival::coxph()} with \texttt{iter.max\ =\ 0}
  evaluates each site's partial log-likelihood at a supplied \(\beta\),
  while \texttt{mle()} performs the optimization.
\item
  \textbf{Symbolic convex programs.} \pkg{CVXR} \citep{CVXRJSS}
  canonicalizes a convex problem rather than repeatedly calling an
  objective. We therefore reformulate the distributed problem as
  consensus ADMM: each site solves its local \pkg{CVXR} subproblem in
  cleartext, and only the cross-site consensus update uses threshold
  FHE.
\end{enumerate}

Shorter sections summarize two additional worked protocols from the
package vignettes: federated similarity retrieval and two-party
encrypted prediction. They illustrate the range of the primitives but
are not validation studies comparable to the Cox and consensus-ADMM
examples. The exact-integer BFV demonstrations are developed only in the
package vignettes.

The \pkg{homomorpheR} package provides the constructs these computations
need. A \texttt{Site} is a participating party; a \texttt{LocalSite},
the kind \texttt{make\_worker()} builds, holds its data and a
\texttt{contribution\_fn}; a \texttt{Master} orchestrates a protocol,
holding a key pair for a single-decrypter \texttt{CKKSMaster} and only
public material for a \texttt{ThresholdMaster}; and
\texttt{master\_aggregate()} drives one round of fan-out and fan-in: it
broadcasts a query \(\theta\), each site returns an encrypted
contribution, and the decrypted sum comes back. In every fit in this
paper \(\theta\) is a model parameter and the runner sits inside an
optimizer's objective function, but a single round is equally a one-shot
aggregate query; in the \texttt{query-count-threshold} vignette
(Section~\ref{sec-discussion}), \(\theta\) is a count predicate. The
package also implements the threshold-key generation ceremony used by
the remote trust model.

The paper is organized as follows. Section~\ref{sec-fhe} introduces the
FHE schemes the packages support, with emphasis on CKKS for real-valued
statistical computation, and closes with the threat model
(Section~\ref{sec-threat-model}) that every later security statement
refers back to. Section~\ref{sec-openfhe} describes the \pkg{openfhe.R}
binding layer and Section~\ref{sec-homomorpheR} the \pkg{homomorpheR}
class hierarchy. The two substantial examples follow:
Section~\ref{sec-cox} fits a distributed stratified Cox regression in
single-decrypter and threshold variants, and Section~\ref{sec-cvxr} a
federated Cox-lasso by consensus ADMM with \pkg{CVXR}.
Section~\ref{sec-similarity} and Section~\ref{sec-threats} summarize
additional retrieval and two-party prediction protocols, including their
output leakage. Section~\ref{sec-dp} records an exploratory composition
with Gaussian noise; it is explicitly not a differential-privacy
guarantee. Section~\ref{sec-related} places the work among its
predecessors, and Section~\ref{sec-discussion} concludes.

\section{FHE in brief}\label{sec-fhe}

Homomorphic encryption has its own vocabulary, and we use only a small
part of it, in terms familiar to a statistician. Where a term has a
standard cryptographic name, we note it within parentheses.

FHE operates on encrypted data through a public-key system with three
roles. A \emph{key generator} produces a public--secret key pair and,
optionally, evaluation keys that authorize specific operations such as
multiplication and rotation of vector elements. A \emph{data holder}
encrypts data under the public key. A \emph{computing party} (the
\emph{evaluator}), such as an aggregator, applies arithmetic or logical
operations to the encrypted values using the evaluation keys and needs
no secret key to do so. When the holder of the secret key decrypts the
result, it is the value the same operations would have produced on the
unencrypted data: exactly for the integer schemes below, and to within a
small approximation error for the real-valued one. The encrypted values
themselves reveal nothing about the data to the computing party.

OpenFHE implements three families of schemes, each suited to a different
type of computation. The \pkg{openfhe.R} package exposes all three.

\subsection{BFV and BGV: exact integer arithmetic}\label{sec-bfv-bgv}

OpenFHE provides the BFV \citep{FanVercauteren2012} and BGV
\citep{BrakerskiVaikuntanathan2014} schemes for exact arithmetic on
integer vectors modulo a prime \(p\). These are useful for counting,
cross-tabulation, and any aggregation of integer-valued summaries where
no approximation error is acceptable. The \pkg{openfhe.R} package
exposes both via \texttt{fhe\_context("BFV",\ ...)} and
\texttt{fhe\_context("BGV",\ ...)}. These find use in aggregation
queries; Section~\ref{sec-discussion} points to the package vignettes
that develop them.

\begin{longtable}[]{@{}
  >{\raggedright\arraybackslash}p{(\linewidth - 2\tabcolsep) * \real{0.4000}}
  >{\raggedright\arraybackslash}p{(\linewidth - 2\tabcolsep) * \real{0.6000}}@{}}
\toprule\noalign{}
\begin{minipage}[b]{\linewidth}\raggedright
Term
\end{minipage} & \begin{minipage}[b]{\linewidth}\raggedright
Meaning
\end{minipage} \\
\midrule\noalign{}
\endfirsthead
\toprule\noalign{}
\begin{minipage}[b]{\linewidth}\raggedright
Term
\end{minipage} & \begin{minipage}[b]{\linewidth}\raggedright
Meaning
\end{minipage} \\
\midrule\noalign{}
\endhead
\bottomrule\noalign{}
\tabularnewline
\caption{Terminology used throughout the
paper.}\label{tbl-terms}\tabularnewline
\endlastfoot
Encrypted value (\emph{ciphertext}) & The result of encrypting a number
or a vector of numbers. Encrypted values can be added and multiplied
without being decrypted first. \\
Cleartext (\emph{plaintext}) & The ordinary, unencrypted number: what
one starts with and what decryption returns. \\
Site & A party holding data that cannot leave the institution that
collected it. \\
Aggregator & The party that collects encrypted per-site contributions
and combines them while they remain encrypted. What it ends up seeing
depends on the protocol and is stated case by case; in the master/worker
fits below it is the total and nothing else. \\
Protocol & A procedure built for a particular problem: what each site
computes, and what the aggregator does with the total. \\
Slot & An encrypted value is a vector of many positions; one arithmetic
operation acts on every position at once. \\
Evaluation keys & Auxiliary keys that \emph{authorize} particular
operations --- multiplication, rotation, summation across slots. They
permit computation, not decryption. \\
Threshold keys & A key arrangement in which no single party holds the
secret key; decryption requires every party to contribute. \\
Multiplicative depth & How many multiplications deep a computation may
go. The parameters fix a chain of moduli when the context is built, and
each multiplication consumes one link. Additions consume none. \\
Precision & How closely a decrypted real-valued result matches the
cleartext computation. Related to depth but not the same thing: noise
accumulates with the operations performed, while depth is a hard limit
on how many multiplications are possible at all. \\
\end{longtable}

\subsection{CKKS: approximate real-number arithmetic}\label{sec-ckks}

The Cheon--Kim--Kim--Song (CKKS) scheme \citep{CheonKimKimSong2017}
encodes vectors of real (or complex) numbers into encrypted values and
supports approximate arithmetic on them. Each homomorphic operation
introduces a small amount of noise; the precision depends on the
encoding parameters and diminishes with each multiplication. How much
error a particular computation incurs therefore depends on that
computation, and we report it where each example is presented rather
than quoting a figure for the scheme.

CKKS is the scheme used in all the statistical demonstrations that
follow. Every real-valued computation --- partial log-likelihoods, score
functions, distributed optimization updates --- is carried out under
CKKS.

The CKKS encoding works as follows. A vector of real values
\((v_1, \ldots, v_n)\) is multiplied by a \emph{scaling factor}
\(\Delta = 2^s\) to produce integer approximations, which are then
encoded into a polynomial. Encryption adds a small noise term to this
polynomial. After a homomorphic addition the scaling factor of the
result is still \(\Delta\) and the decrypted values are the element-wise
sums of the inputs, perturbed by the accumulated noise. After a
homomorphic multiplication, however, the scaling factor of the product
is \(\Delta^2\). A \emph{rescaling} step divides the encrypted value by
\(\Delta\), returning the scaling factor to \(\Delta\) and restoring the
original precision --- but at the cost of consuming one level of the
chain of moduli.

That chain of prime moduli is fixed when the context is built; the user
sets its length via \texttt{multiplicative\_depth}. A depth of \(d\)
allows \(d\) sequential multiplications interspersed with rescales
before the chain is exhausted and further multiplications return
detectably wrong values. The aggregation patterns in
Section~\ref{sec-cox} and Section~\ref{sec-cvxr} require depth 1; the
bootstrapping primitives in the package's \texttt{ckks-bootstrapping}
vignette refresh the noise and restore the full budget at the cost of
additional computation per refresh. A CKKS encrypted value is also a
vector: it holds up to \texttt{batch\_size} real-valued slots, and a
single operation acts on every slot at once.

A typical CKKS context for these demonstrations is

\begin{verbatim}
cc <- openfhe.R::fhe_context(scheme               = "CKKS",
                             multiplicative_depth = 1L,
                             scaling_mod_size     = 50L,
                             batch_size           = 8L)
\end{verbatim}

with the security level defaulting to \texttt{HEStd\_128\_classic}
(128-bit classical security, the current standard recommendation for
applied FHE). The OpenFHE documentation discusses scaling techniques,
key-switching algorithms, and parameter trade-offs; we list only the
parameters needed for the demonstrations in this paper.

\subsection{FHEW and TFHE: Boolean operations}\label{sec-binfhe}

OpenFHE also provides the FHEW \citep{DucasMicciancio2015} and TFHE
\citep{ChillottiGamaGeorgievaIzabachene2020} schemes for Boolean gate
evaluation on encrypted bits, with programmable bootstrapping that
refreshes the noise after every gate so that the depth is effectively
unlimited. The \pkg{openfhe.R} package exposes these via
\texttt{bin\_fhe\_context()} and the \texttt{binfhe-boolean-circuits}
vignette. The statistical demonstrations in this paper do not use them,
since the computations they perform are real-valued and handled by CKKS.

\subsection{Multiparty extensions}\label{sec-multiparty}

A standard CKKS or BFV deployment leaves a residual trust issue: some
single party holds the secret key. OpenFHE's \emph{multiparty} extension
\citep{MouchetTroncosoPastorizaBossuatHubaux2021} removes this
assumption. The library's \texttt{MultipartyKeyGen()}, wrapped in
\pkg{openfhe.R} as \texttt{multiparty\_key\_gen()}, extends a public key
party by party, so that the \emph{joint} public key is built from all
parties' contributions, and each party retains only its own secret
share. No proper subset of shares can decrypt; decryption requires every
party to contribute a partial decryption, which the aggregator then
fuses. We use this construction in Section~\ref{sec-cox-threshold} and
Section~\ref{sec-cvxr}; it is the main reason \pkg{homomorpheR} ships a
dedicated \texttt{ThresholdMaster} class.

\subsection{Threat model and scope}\label{sec-threat-model}

Every security statement later in the paper is relative to the
assumptions set out here, and they are narrower than the phrase ``the
data are encrypted'' suggests.

\begin{enumerate}
\def\labelenumi{\arabic{enumi}.}
\item
  \emph{Honest-but-curious parties.} The parties follow the protocol and
  may try to learn from what they legitimately observe. We do not detect
  deviations: a site can return a well-formed encrypted value that is
  not its honest contribution, or a malformed partial decryption that
  corrupts a fused result with no error raised anywhere. Defending
  against such possibilities needs verifiable decryption and committed
  key generation, which these packages do not provide.
\item
  \emph{Who can decrypt.} With a single decrypter, the aggregator holds
  the secret key. Each site still returns an encrypted value, but the
  aggregator could decrypt any individual contribution; ``only the
  aggregate is decrypted'' describes what the runner does, not something
  the cryptography prevents. With \emph{threshold} keys
  (Section~\ref{sec-multiparty}), no party holds the secret key, and
  decryption needs a partial decryption from every site, so no single
  party can decrypt anything on its own.
\item
  \emph{Where the parties run.} In a single session, every party is an
  object in one \proglang{R} process. The \texttt{Site} and
  \texttt{Master} classes model the protocol's roles; they create no
  process or trust boundary, and one object can reach another. A
  threshold master has no secret-key or secret-share property and its
  methods use no secret material, but the site objects it queries live
  in the same process, so their shares are reachable from its object
  graph. With separate endpoints, each site is a separately controlled
  process that generates and keeps its own share and returns only
  encrypted values or partial decryptions. Only then does a party
  boundary exist, and only if the deployment's transport,
  authentication, endpoint code, and key storage establish one.
\item
  \emph{What the protocol reveals by design.} Encryption protects the
  channel and the intermediate computation; the protocol decides what is
  revealed. Three such disclosures recur below. The decrypted output of
  each round is by construction visible to whoever receives it, and a
  party choosing query points adaptively can often invert the function
  being evaluated --- Section~\ref{sec-threats} demonstrates this
  against a linear model. Next, in the runner of Figure
  \ref{fig-aggregate}, we allow missingness to travel in the clear,
  because CKKS has no representation for \texttt{NA}, so an aggregator
  learns which parameter values a site could not evaluate. (Other
  protocols may handle this differently.) Finally, the retrieval
  protocol of Section~\ref{sec-similarity} decrypts every candidate
  score in order to rank them, which is more than the top-\(k\) it
  returns.
\end{enumerate}

Every example and vignette, under both key arrangements, runs in a
single session, so that a reader can reproduce the results. Separate
endpoints are specified by the \texttt{RemoteSite} contract, and no
example runs them, because \pkg{homomorpheR} ships no transport.

A \texttt{RemoteSite} subclass implements two operations, corresponding
to the only two moments anything passes between the parties: a setup
call that provisions the endpoint with the public context and key, and a
per-round call that sends the query parameter and returns a value the
endpoint has already encrypted. Threshold protocols add two more, a key
generation round that retains the new share remotely and a partial
decryption that returns only a partial. What the package does enforce is
narrow: before adding a reply to the total, \texttt{master\_aggregate()}
refuses one that is not an encrypted value, or that was produced under a
different key, which it detects from the key tag OpenFHE stamps on every
encrypted value, using public data only. Transport security, identity,
authentication, attestation, and remote key storage belong to the
deployment and so are beyond our scope.

\section{The openfhe.R package}\label{sec-openfhe}

The \pkg{openfhe.R} package presents two user-facing layers: an \pkg{S7}
\citep{S7} class hierarchy that wraps the underlying OpenFHE C++ objects
using \pkg{cpp11} \citep{cpp11}, and an \proglang{R} operator-dispatch
layer that makes arithmetic on encrypted data look like ordinary
arithmetic.

\subsection{The S7 class hierarchy}\label{sec-openfhe-s7}

Each OpenFHE object type is represented by an \pkg{S7} class. The main
ones are \texttt{CryptoContext}, \texttt{PublicKey},
\texttt{PrivateKey}, \texttt{Plaintext}, \texttt{Ciphertext},
\texttt{EvalKey}, and (for binary FHE) \texttt{BinFHEContext},
\texttt{LWECiphertext}, and \texttt{LWEPrivateKey}; the others include
the parameter classes \texttt{CKKSParams}, \texttt{BFVParams}, and
\texttt{BGVParams}. All inherit from a base class \texttt{OpenFHEObject}
that backs the underlying C++ object. The one exception is
\texttt{KeyPair}, a compound \proglang{R}-side container with no C++
object of its own and two properties, \texttt{public} and
\texttt{secret}, holding the public and private keys respectively; the
user accesses them with the \pkg{S7} \texttt{@} operator.

OpenFHE's \texttt{Feature} enum controls which operations are available
on a given context. The three features enabled by default ---
\texttt{PKE}, \texttt{KEYSWITCH}, and \texttt{LEVELEDSHE} --- cover
encryption, key switching, and the standard homomorphic evaluations
(add, subtract, multiply, negate, rotate). Additional features such as
\texttt{MULTIPARTY} (for threshold protocols), \texttt{ADVANCEDSHE} (for
polynomial and Chebyshev evaluation), and \texttt{FHE} (for
bootstrapping) are enabled via the \texttt{features} argument to
\texttt{fhe\_context()}.

\subsection{Binding boundary and
scope}\label{binding-boundary-and-scope}

The S7 objects hold \texttt{cpp11} external pointers to the
corresponding OpenFHE C++ handles; they do not copy encrypted values
into R vectors. Registered C++ entry points unwrap those pointers,
invoke the OpenFHE operation, and wrap the returned handle. A common
exception adapter converts C++ failures at this boundary into R errors
with the operation name, rather than allowing an exception to cross the
C interface. Serialization functions write contexts, keys, and encrypted
values to files, in binary or JSON form, and read them back, but do not
themselves move those files or authenticate their source.

The binding is designed and maintained to mirror the
\pkg{openfhe-python} surface. This paper exercises only the subset
needed by the worked protocols: context and key construction, packed
plaintexts, encryption and decryption, arithmetic, rotations, Chebyshev
evaluation, serialization, and multiparty key generation and decryption.
The examples do not isolate wrapper overhead from cryptographic cost;
they describe complete R-level protocol runs, not a binding
microbenchmark.

\subsection{Operator dispatch}\label{sec-openfhe-ops}

Arithmetic on encrypted values uses \proglang{R}'s \texttt{Ops} group
generic so that the binary operators \texttt{+}, \texttt{-}, \texttt{*}
and the unary \texttt{-} dispatch to the corresponding
\texttt{eval\_add}, \texttt{eval\_sub}, \texttt{eval\_mult}, and
\texttt{eval\_negate} methods. The operands may be two encrypted
vectors, an encrypted vector followed by an unencrypted one, or an
encrypted vector and a numeric scalar in either order; the dispatcher
resolves the correct OpenFHE overload.

The end-to-end pattern of an encrypt--compute--decrypt pipeline under
\pkg{openfhe.R} is short.

\begin{verbatim}
library(openfhe.R)
cc   <- fhe_context(scheme               = "CKKS",
                    multiplicative_depth = 1L,
                    scaling_mod_size     = 50L,
                    batch_size           = 8L)
keys <- key_gen(cc = cc, eval_mult = TRUE)

pt_a <- make_ckks_packed_plaintext(cc = cc, values = c(1.5, 2.0))
pt_b <- make_ckks_packed_plaintext(cc = cc, values = c(3.0, 4.0))
ct_a <- encrypt(key = keys@public, pt = pt_a, cc = cc)
ct_b <- encrypt(key = keys@public, pt = pt_b, cc = cc)

ct_sum <- ct_a + ct_b           # homomorphic addition
ct_dot <- ct_a * 2 + ct_b * 3   # encrypted * cleartext scalar
pt     <- decrypt(ct = ct_sum, key = keys@secret, cc = cc)
get_real_packed_value(pt)[1:2]
\end{verbatim}

A word on names. We use statistical vocabulary wherever we can ---
encrypted values rather than ciphertexts, computing parties rather than
evaluators --- but the argument names in code are not translated.
\texttt{key}, \texttt{pt}, and \texttt{ct} abbreviate the parameters of
the underlying OpenFHE C++ signatures,
\texttt{Encrypt(publicKey,\ plaintext)} and
\texttt{Decrypt(ciphertext,\ privateKey)}, in the same order, and
\texttt{cc} is the crypto context on which C++ calls those methods,
passed in R as a trailing argument; \pkg{openfhe-python} spells the same
parameters out as \texttt{publicKey}, \texttt{plaintext},
\texttt{ciphertext}, and \texttt{privateKey} on methods of the crypto
context. Keeping the order lets a reader move between the C++
documentation, \pkg{openfhe-python}, and \pkg{openfhe.R} with only the
abbreviations to learn, and it is the choice the rest of the stack
inherits, since S7 requires every method on a generic to use the
generic's names for the arguments it dispatches on.

That is the minimal encrypt--compute--decrypt path, and it is the part
of the interface the protocols in this paper rest on. The package API is
considerably richer --- serialization, rotation and automorphism keys,
bootstrapping, Chebyshev evaluation of transcendental functions, and the
Boolean contexts --- and the vignettes illustrate their use, as do
additional examples in Section~\ref{sec-similarity} and
Section~\ref{sec-threats}.

\section{The homomorpheR class hierarchy}\label{sec-homomorpheR}

The \pkg{openfhe.R} package gives the user the cryptographic primitives:
contexts, keys, encrypt, decrypt, homomorphic arithmetic. But the
structural machinery that a multi-party protocol needs is provided by
\pkg{homomorpheR}: sites that hold local data, a master that
orchestrates the round, and generic methods that treat encryption and
decryption as a backend choice rather than something every protocol
re-implements.

The package's class hierarchy is a small set of abstract \pkg{S7} bases
--- one for each role a protocol names --- with concrete subclasses, and
a handful of generics. Every object lives in the session that runs a
protocol, including the one standing in for a party that does not: what
distinguishes a remote site is where its contribution is
\emph{computed}, not where its object lives.

\begin{itemize}
\tightlist
\item
  \texttt{Site} --- abstract base carrying \texttt{name} and a mutable
  \texttt{state} environment, with two branches. \texttt{LocalSite} is
  concrete: it holds the site's \texttt{data} and a reusable
  \texttt{contribution\_fn(data,\ theta)}, and \texttt{contribute()}
  passes the two to that function and encrypts what it returns.
  \texttt{RemoteSite} is itself abstract and adds no properties of its
  own. The package supplies the contract --- four methods that refuse to
  act locally on a remote party's behalf --- and the user supplies the
  subclass that carries them out over a transport, so the contribution
  is computed and encrypted at the far end. Both branches answer the
  same generic, \texttt{contribute(site,\ theta)}, with an encrypted
  contribution.
\item
  \texttt{Master} --- abstract base carrying the same \texttt{name} and
  \texttt{state}. Its subclasses built on \pkg{openfhe.R} are
  \texttt{CKKSMaster} (single-decrypter CKKS) and
  \texttt{ThresholdMaster} (n-of-n threshold key generation, no single
  decrypter). Both are constructed and driven in the same session as the
  sites they query.
\end{itemize}

A user develops a protocol for a particular problem: what each site
computes, and what the aggregator does with the total. Four function
calls help in the construction: one is made at wiring time, by
\texttt{set\_workers()} or \texttt{make\_threshold\_master()}, and
configures a site; two are made by the runner --- one queries a site,
one recovers the total --- and the fourth is made by a party on its own
behalf:

\begin{verbatim}
## At wiring time, on each site:
set_public_params(site, params)          # setup: give a site its parameters

## The runner, on a site and on the aggregate:
contribute(site, theta)                  # -> encrypted contribution, or NA
decrypt(master, ciphertext, len)         # -> numeric

## A party, on its own value, with no counterparty:
encrypt(site, value)                     # -> encrypted value
\end{verbatim}

\texttt{encrypt()} is the single encryption entry point, and the two
uses nest: \texttt{contribute()} calls it for a \texttt{LocalSite} whose
\texttt{contribution\_fn} returned a plain number, while a site calls it
directly whenever its contribution is not a single number, as the
per-site moments and consensus terms of Section~\ref{sec-cvxr} each do.
Together with \texttt{decrypt()} it abstracts the cryptographic backend,
so neither the runner nor a site depends on which scheme is in use.

Both generics belong to \pkg{openfhe.R}; \pkg{homomorpheR} registers its
methods on them and re-exports them, so there is one \texttt{encrypt()}
and one \texttt{decrypt()} across the two packages rather than a
same-named pair that would shadow each other.

A site's \texttt{contribution\_fn} produces a plain number, and that
number never leaves the site: \texttt{contribute()} encrypts it there,
with the public parameters the site was given at setup, so what the
runner receives is already encrypted. This is the ordering the whole
construction depends on, and \texttt{master\_aggregate()} enforces it by
rejecting any reply that is not an encrypted value under the expected
key.

\begin{figure}[!ht]
  \centering
  \resizebox{\linewidth}{!}{\begin{tikzpicture}[
    node distance = 14mm and 16mm,
    site/.style   = {draw, rounded corners, thick, minimum width = 20mm,
                     minimum height = 9mm, align = center, fill = blue!8},
    master/.style = {draw, rounded corners, thick, minimum width = 26mm,
                     minimum height = 11mm, align = center, fill = orange!15},
    arr/.style    = {-{Latex[length=2.0mm]}, thick},
    label/.style  = {font = \small\itshape, color = gray!60!black},
    every node/.append style = {font = \small}
  ]
  \begin{scope}
    \node[master] (M1)  at (0, 0)    {Master\\\scriptsize{(holds $sk$)}};
    \node[site]   (W1)  at (-3.0, -3.0) {Site 1\\\scriptsize{$pk$}};
    \node[site]   (W2)  at ( 0,   -5.0) {Site 2\\\scriptsize{$pk$}};
    \node[site]   (W3)  at ( 3.0, -3.0) {Site 3\\\scriptsize{$pk$}};

    \draw[arr] (M1) to[bend right=14]
      node[midway, sloped, above, label] {$\theta$} (W1);
    \draw[arr] (W1) to[bend right=14]
      node[midway, sloped, above, label] {$E(\ell_1)$} (M1);

    \draw[arr] (M1) to[bend right=12]
      node[midway, sloped, below, label] {$\theta$} (W2);
    \draw[arr] (W2) to[bend right=12]
      node[midway, sloped, below, label] {$E(\ell_2)$} (M1);

    \draw[arr] (M1) to[bend left=14]
      node[midway, sloped, above, label] {$\theta$} (W3);
    \draw[arr] (W3) to[bend left=14]
      node[midway, sloped, above, label] {$E(\ell_3)$} (M1);

    \node[align = center, font = \small\bfseries] at (0, -6.4)
      {Single-decrypter master/worker};
    \node[align = center, font = \scriptsize, color = gray!60!black] at (0, -6.9)
      {master decrypts $\boxplus_i E(\ell_i) = E(\sum_i \ell_i)$};
  \end{scope}

  \begin{scope}[xshift = 9.0cm]
    \node[master] (M2)  at (0, 0)    {Aggregator\\\scriptsize{(no $sk$)}};
    \node[site]   (T1)  at (-3.0, -3.0) {Site 1\\\scriptsize{$sk_1,\,pk_{1..N}$}};
    \node[site]   (T2)  at ( 0,   -5.0) {Site 2\\\scriptsize{$sk_2,\,pk_{1..N}$}};
    \node[site]   (T3)  at ( 3.0, -3.0) {Site 3\\\scriptsize{$sk_3,\,pk_{1..N}$}};

    \draw[arr] (M2) to[bend right=14]
      node[midway, sloped, above, label] {$\theta$} (T1);
    \draw[arr] (T1) to[bend right=14]
      node[midway, sloped, above, label] {$E(\ell_1)$} (M2);

    \draw[arr] (M2) to[bend right=12]
      node[midway, sloped, below, label] {$\theta$} (T2);
    \draw[arr] (T2) to[bend right=12]
      node[midway, sloped, below, label] {$E(\ell_2)$} (M2);

    \draw[arr] (M2) to[bend left=14]
      node[midway, sloped, above, label] {$\theta$} (T3);
    \draw[arr] (T3) to[bend left=14]
      node[midway, sloped, above, label] {$E(\ell_3)$} (M2);

    \node[align = center, font = \small\bfseries] at (0, -6.4)
      {Threshold master/worker};
    \node[align = center, font = \scriptsize, color = gray!60!black] at (0, -6.9)
      {decryption is $n$-of-$n$: each site contributes a partial};
  \end{scope}
\end{tikzpicture}}
  \caption{The master/worker topology in single-decrypter (left)
  and threshold (right) variants. Each site was given the
  encryption key $pk$ (or the joint $pk_{1..N}$ in the threshold
  case) once, at setup; a round broadcasts only the
  parameter $\theta$. Each site $i$
  returns an encrypted local summary $E(\ell_i)$; the encrypted
  contributions are summed homomorphically. The single-decrypter
  master holds the secret key $sk$ alone and decrypts the total
  directly. In the threshold case, each site $i$ holds only its
  own secret share $sk_i$ and contributes a partial decryption
  that the aggregator fuses ($n$-of-$n$); no single party can
  decrypt. The \texttt{master\_aggregate()} runner body is the
  same in both cases; only the master's class differs.}
  \label{fig-mw}
\end{figure}
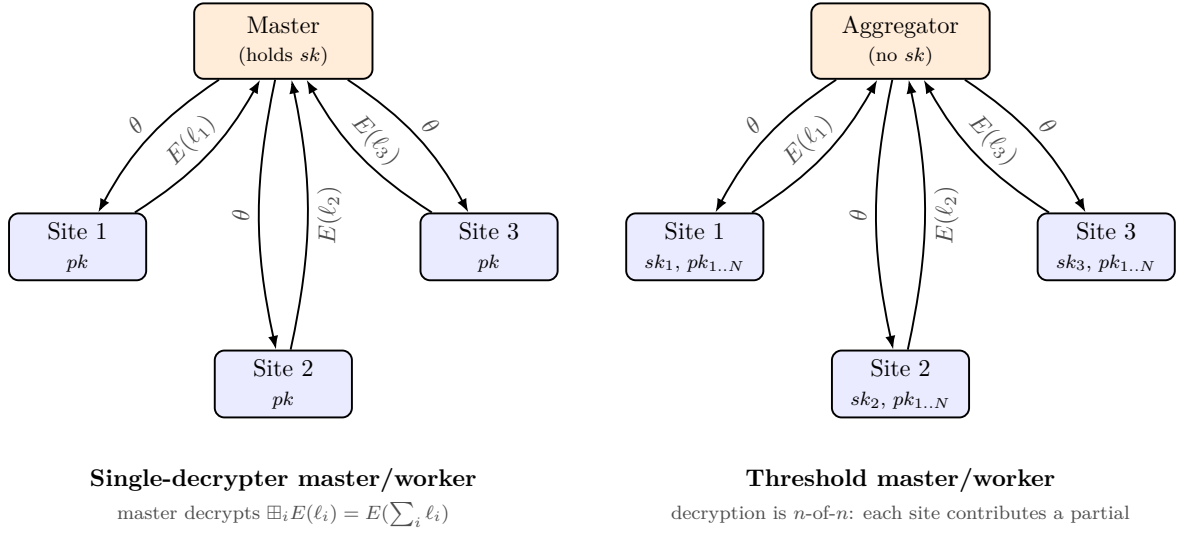

The two setup and encryption calls take no aggregator argument at all,
and the asymmetry with \texttt{decrypt()}, which takes one, is important
to note. Decryption is privileged: it needs secret material, or, under
threshold keys, the standing to convene every site. Encryption and setup
need only public material. A site is autonomous once configured --- it
is handed its parameters once, and from then on computes and encrypts
without consulting anyone. For a remote site that setup call is a
message its implementation must deliver to the far endpoint per
Section~\ref{sec-threat-model}. Figure \ref{fig-mw} shows the shape of
one round under both trust models.

The site interface has local and remote implementations that share
method dispatch but not deployment semantics. \texttt{make\_worker()}
builds a \texttt{LocalSite}, whose records and \texttt{contribution\_fn}
are in the current R process. \texttt{RemoteSite} is an abstract
integration point for a separately controlled endpoint.
\pkg{homomorpheR} supplies no concrete remote transport: users must
subclass it and implement the setup and \texttt{contribute()} contract
described in Section~\ref{sec-threat-model}. A correct implementation
encrypts at that endpoint; the aggregator never receives the
contribution in cleartext.

A site's \texttt{contribute()} can fail for reasons that have nothing to
do with the statistics, such as a network failure or an unreachable
remote endpoint. It should then call \texttt{site\_unavailable()}, which
aborts the round. It should not return \texttt{NA} or a stand-in such as
an encrypted zero. \texttt{NA} would tell the optimizer that \(\theta\)
is infeasible, and an encrypted zero would be summed as if it were the
site's value; either way the fit goes wrong with no error raised.
\texttt{NA} is reserved for the case described below, a site that cannot
evaluate its contribution at \(\theta\).

Throughout, \emph{worker} is simply our name for a \texttt{Site} that
has been wired to a master: \texttt{make\_worker()} constructs a
\texttt{Site}, \texttt{set\_workers()} attaches a list of them to a
master, and the master fans \(\theta\) out to its workers and aggregates
their encrypted contributions.

\begin{figure}[t]
  \centering
  \fbox{%
    \begin{minipage}{\dimexpr\linewidth-2\fboxsep-2\fboxrule\relax}
      \begin{algorithmic}
        \Function{master\_aggregate}{$\mathit{master},\ \theta$}
          \For{each worker $i$ wired to the master}
            \State $c_i \gets \mathrm{contribute}(\mathit{worker}_i,\ \theta)$
              \Comment{worker $i$ encrypts; master never sees $\ell_i$}
            \If{$c_i$ is \texttt{NA}}
              \State \Return \texttt{NA} \Comment{$\theta$ non-evaluable there}
            \EndIf
          \EndFor
          \State $c \gets \sum_i c_i$ \Comment{homomorphic summation}
          \State \Return $\mathrm{decrypt}(\mathit{master},\ c)$
        \EndFunction
      \end{algorithmic}
    \end{minipage}%
  }
  \caption{The exported \pkg{homomorpheR} runner
  \texttt{master\_aggregate()} --- one round of encrypted fan-out /
  fan-in at query $\theta$. A \emph{worker} is a
  \texttt{Site} that \texttt{set\_workers()} has wired to this
  master. Only $\theta$ is broadcast, and each worker encrypts its own
  contribution $\ell_i$ under the public key it was given to return an
  encrypted value. \texttt{NA} is the sole cleartext exception, for the
  reason given in Section~\ref{sec-threat-model}.}
  \label{fig-aggregate}
\end{figure}

Figure \ref{fig-aggregate} gives the runner itself. Its body is the same
for both master variants and both site classes; dispatch selects
\texttt{contribute()} for the site and \texttt{decrypt()} for the
master. The encrypted-reply and key-identity checks described in
Section~\ref{sec-threat-model} apply before aggregation. A site whose
objective is undefined at \(\theta\) returns \texttt{NA} in the clear,
and the runner hands it back to its caller; an optimizer, the caller in
every fit below, backs off as it would on cleartext data. Under
threshold keys a site can instead have its contribution function return
a large finite value there, as objectives handed to \texttt{optim()}
sometimes do, so that the aggregator does not learn which site it was.
The matching constructors and helpers are listed in
Table~\ref{tbl-mw-api}.

\begin{longtable}[]{@{}
  >{\raggedright\arraybackslash}p{(\linewidth - 2\tabcolsep) * \real{0.5000}}
  >{\raggedright\arraybackslash}p{(\linewidth - 2\tabcolsep) * \real{0.5000}}@{}}
\toprule\noalign{}
\begin{minipage}[b]{\linewidth}\raggedright
Constructor / helper
\end{minipage} & \begin{minipage}[b]{\linewidth}\raggedright
Purpose
\end{minipage} \\
\midrule\noalign{}
\endfirsthead
\toprule\noalign{}
\begin{minipage}[b]{\linewidth}\raggedright
Constructor / helper
\end{minipage} & \begin{minipage}[b]{\linewidth}\raggedright
Purpose
\end{minipage} \\
\midrule\noalign{}
\endhead
\bottomrule\noalign{}
\tabularnewline
\caption{\pkg{homomorpheR} master/worker constructors and
helpers.}\label{tbl-mw-api}\tabularnewline
\endlastfoot
\texttt{make\_worker(name,\ data,\ contribution\_fn)} & Build a worker
--- a \texttt{LocalSite} from one site's data and its
\texttt{contribution\_fn} \\
\texttt{contribute(site,\ theta)} & A site's \emph{encrypted}
contribution at \texttt{theta}; the one call the runner makes on a
site \\
\texttt{site\_unavailable(message,\ site\ =\ NULL,\ parent\ =\ NULL)} &
Signal that a site could not be reached --- distinct from an \texttt{NA}
contribution \\
\texttt{make\_ckks\_master(name,\ crypto\_context,\ keypair)} &
Single-decrypter CKKS master \\
\texttt{make\_threshold\_master(name,\ crypto\_context,\ sites)} &
Threshold master; runs the key ceremony through \texttt{sites}, each
keeping its own share, and returns the master wired to them \\
\texttt{set\_workers(master,\ workers)} & Wire a list of workers to a
master (single-decrypter masters) \\
\texttt{master\_aggregate(master,\ theta)} & One round: fan \(\theta\)
out, aggregate the encrypted contributions, decrypt \\
\end{longtable}

The example below provides an illustrative use of the constructors in a
distributed Poisson maximum likelihood estimation task across three
sites using CKKS. The data, the per-site negative log-likelihood, and
the optimizer are easily recognizable to any \proglang{R} programmer;
the cryptographic setup is four calls --- a context, a key pair, a
master, and the wiring --- plus one \texttt{master\_aggregate()} inside
the objective. Arguments are named here for the same reason as in
Section~\ref{sec-openfhe}: the wiring should be readable without
consulting a signature.

\begin{verbatim}
set.seed(17822)
y <- rpois(40, lambda = 10)
y1 <- y[1:20]
y2 <- y[21:27]
y3 <- y[28:40]

local_nll <- function(data, lambda) {
    -sum(stats::dpois(data, lambda = lambda, log = TRUE))
}

cc   <- openfhe.R::fhe_context(scheme               = "CKKS",
                               multiplicative_depth = 1L,
                               scaling_mod_size     = 50L,
                               batch_size           = 8L)
keys <- openfhe.R::key_gen(cc = cc)

master <- make_ckks_master(name           = "Master",
                           crypto_context = cc,
                           keypair        = keys)
set_workers(master  = master,
            workers = list(
                make_worker(name = "S1", data = y1, contribution_fn = local_nll),
                make_worker(name = "S2", data = y2, contribution_fn = local_nll),
                make_worker(name = "S3", data = y3, contribution_fn = local_nll)))

fit <- stats4::mle(
    minuslogl = function(lambda) {
        master_aggregate(master = master, theta = lambda)
    },
    start = list(lambda = 5))
\end{verbatim}

The recovered MLE is \(\hat\lambda = 9.175\), and the cleartext sample
mean is \(\bar y = 9.175\); they differ by 4.0e-07. Across the run,
every per-site negative log-likelihood traveled to the master encrypted
under the master's public key; only the \emph{aggregate} negative
log-likelihood was decrypted at each \texttt{mle()} query.

\section{Distributed Cox regression}\label{sec-cox}

The Cox proportional-hazards model \citep{Cox1972} is one of the
workhorses of medical statistics. Given covariates \(x_i\) for subject
\(i\), an event time \(t_i\), and an event indicator \(\delta_i\), the
hazard is modeled as

\[
h(t \mid x_i) \;=\; h_0(t) \, \exp(\beta^\top x_i),
\]

with an unspecified baseline hazard \(h_0(t)\) and a vector \(\beta\) of
regression coefficients. Estimation is by maximization of the partial
log-likelihood

\[
\ell(\beta) \;=\; \sum_{i: \delta_i = 1}
  \!\left[\, \beta^\top x_i \;-\; \log\!\!\sum_{j \in R_i} \exp(\beta^\top x_j) \,\right],
\]

where \(R_i\) is the risk set at time \(t_i\)
\citep{TherneauGrambsch2000}.

For \emph{stratified} Cox regression the partial log-likelihood
factorizes additively across strata,
\(\ell(\beta) = \sum_s \ell_s(\beta)\). When the strata are the sites,
this is the additive decomposition the \pkg{homomorpheR} master/worker
classes need: each site computes its own \(\ell_s(\beta)\) on its
private data; the master sums the encrypted contributions and decrypts
the total. The mathematical decomposition is identical to the cleartext
case --- the cryptographic protocol replaces only the \emph{transport}
of intermediate sums.

\subsection{DLBCL lymphoma cohort}\label{dlbcl-lymphoma-cohort}

We use the diffuse large-B-cell lymphoma (DLBCL) gene-expression cohort
of \citet{Rosenwald2002} --- the same data that \citet{BayleFanLou2025}
use for distributed Cox estimation. Patients are matched to the
Lymphochip expression columns by their LYM number; of the 7399
microarray features we retain the 6416 observed in at least 75\% of the
240 patients, imputing their remaining missing values by the per-feature
mean; and, following \citet{BayleFanLou2025}, we exclude the five
patients with zero follow-up time. That leaves \(n = 235\) patients with
133 deaths over a median follow-up of 2.8 years. The processed clinical
table and the full expression matrix are distributed with
\pkg{homomorpheR} as \texttt{DLBCL} and \texttt{DLBCL\_gex}.

The published outcome predictor combines four gene-expression signatures
(germinal-center B cell, lymph node, proliferation, and MHC class II)
and the single gene BMP6. We model the hazard as a function of those
five variables, stratified by molecular subgroup --- GCB, ABC, and Type
III --- and \emph{use the subgroup itself as the site boundary}, giving
three imbalanced sites: GCB \(n = 115\), ABC \(n = 71\), Type III
\(n = 49\), with 54, 49, and 30 deaths respectively. Expression values
are standardized inside the encrypted pipeline of
Section~\ref{sec-cvxr}. The four signature values are averages of
median-centered log-ratio expression values over the genes of each
signature, and BMP6 is the median-centered log ratio of one gene
\citep{Rosenwald2002}; all five enter the model as published, without
further scaling.

Using the molecular subgroup as the site boundary is a choice made only
to demonstrate our example.

The stratified partial log-likelihood factors as
\(\ell(\beta) = \sum_s \ell_s(\beta)\), summed over the three strata
(GCB, ABC, and Type III), so the encrypted-channel master/worker fit
targets exactly the centralized stratified \texttt{coxph} objective.
Table~\ref{tbl-cox-cohort} summarizes the per-subgroup cohort and the
corresponding aggregated fit from the code below.

\begin{verbatim}
agg_model <- coxph(Surv(time, status) ~ GCB_sig + LN_sig +
                       Prolif_sig + BMP6 + MHC2_sig +
                       strata(Subgroup),
                   data = DLBCL)
\end{verbatim}

\begin{table}

\begin{minipage}[t]{0.50\linewidth}

\centering{

\begin{tabular}{lrr}
\toprule
Subgroup & n & Events\\
\midrule
GCB & 115 & 54\\
ABC & 71 & 49\\
Type III & 49 & 30\\
\bottomrule
\end{tabular}

}

\subcaption{\label{tbl-cox-cohort-1}Per-subgroup sample sizes and events.}

\end{minipage}%
\begin{minipage}[t]{0.50\linewidth}

\centering{

\begin{tabular}{lrr}
\toprule
Signature & \(\hat\beta\) & HR\\
\midrule
GCB\_sig & -0.2639 & 0.7681\\
LN\_sig & -0.2544 & 0.7754\\
Prolif\_sig & 0.3031 & 1.3541\\
BMP6 & 0.3036 & 1.3548\\
MHC2\_sig & -0.3191 & 0.7268\\
\bottomrule
\end{tabular}

}

\subcaption{\label{tbl-cox-cohort-2}Centralized stratified coxph() coefficients on the published log-ratio signature scale.}

\end{minipage}%

\caption{\label{tbl-cox-cohort}DLBCL cohort and centralized stratified
Cox fit on the five published expression variables.}

\end{table}%

\subsection{The local Cox partial
log-likelihood}\label{the-local-cox-partial-log-likelihood}

The per-site \(\mathit{contribution\_fn}_i\) of Figure
\ref{fig-aggregate} --- the \texttt{contribution\_fn} argument of
\texttt{make\_worker()} --- is here \texttt{local\_cox\_nll}, shown
below. It exploits the \texttt{iter.max\ =\ 0} feature of the
\texttt{control} parameter of \texttt{survival::coxph()}, which makes
the fit evaluate the partial log-likelihood at \texttt{init} and stop
rather than iterate. The optimization is \texttt{mle()}'s; the
likelihood is \texttt{coxph()}'s.

\begin{verbatim}
local_cox_nll <- function(data, beta) {
    fit <- tryCatch(
        coxph(Surv(time, status) ~ GCB_sig + LN_sig + Prolif_sig +
                  BMP6 + MHC2_sig,
              data    = data,
              init    = beta,
              control = coxph.control(iter.max = 0)),
        error = function(e) NULL)
    if (is.null(fit)) NA_real_ else -fit$loglik[1]
}
\end{verbatim}

The per-site log-likelihoods sum to the stratified one: at \(\beta = 0\)
the three sites sum to 516.598551 against the centralized 516.598551,
agreeing to \ensuremath{2\times 10^{-13}}, giving us a sense of the
accuracy the encrypted channel then has to preserve.

\subsection{Single-decrypter master/worker
fit}\label{single-decrypter-masterworker-fit}

The summed stratified Cox nLL on this cohort is 517 at \(\beta = 0\). We
lift \texttt{scaling\_mod\_size} from 50 to 59 for extra precision and
state \texttt{first\_mod\_size\ =\ 60} explicitly, which is the library
default.

\begin{verbatim}
cc <- openfhe.R::fhe_context(scheme               = "CKKS",
                             multiplicative_depth = 1L,
                             scaling_mod_size     = 59L,
                             first_mod_size       = 60L,
                             batch_size           = 8L)
keys <- openfhe.R::key_gen(cc = cc)

workers <- Map(function(nm, d)
                   make_worker(name            = nm,
                               data            = d,
                               contribution_fn = local_cox_nll),
               names(cox_data), cox_data)

master <- make_ckks_master(name           = "Master",
                           crypto_context = cc,
                           keypair        = keys)
set_workers(master = master, workers = workers)

fit <- stats4::mle(
  minuslogl = function(GCB_sig, LN_sig, Prolif_sig, BMP6, MHC2_sig)
      master_aggregate(master = master,
                       theta  = c(GCB_sig, LN_sig, Prolif_sig,
                                  BMP6, MHC2_sig)),
  start     = list(GCB_sig = 0, LN_sig = 0, Prolif_sig = 0,
                   BMP6    = 0, MHC2_sig = 0),
  method    = "BFGS",
  control   = list(reltol = 1e-7))
\end{verbatim}

The encrypted fits here can be computed from scratch if needed, but for
convenience they are saved in \pkg{homomorpheR} as
\texttt{data(cox\_results)} and \texttt{data(cox\_threshold\_results)}.
To check the encrypted fit, we run the identical \texttt{mle()}
objective with the encrypted aggregation replaced by an ordinary sum.

\begin{verbatim}
plain_aggregate <- function(theta)
    sum(vapply(cox_data, local_cox_nll, numeric(1), beta = theta))

fit_plain <- stats4::mle(
  minuslogl = function(GCB_sig, LN_sig, Prolif_sig, BMP6, MHC2_sig)
      plain_aggregate(c(GCB_sig, LN_sig, Prolif_sig, BMP6, MHC2_sig)),
  start   = list(GCB_sig = 0, LN_sig = 0, Prolif_sig = 0,
                 BMP6    = 0, MHC2_sig = 0),
  method  = "BFGS",
  control = list(reltol = 1e-7))
\end{verbatim}

{

\begin{longtable}[]{@{}
  >{\raggedright\arraybackslash}p{(\linewidth - 6\tabcolsep) * \real{0.1739}}
  >{\raggedleft\arraybackslash}p{(\linewidth - 6\tabcolsep) * \real{0.2464}}
  >{\raggedleft\arraybackslash}p{(\linewidth - 6\tabcolsep) * \real{0.2464}}
  >{\raggedright\arraybackslash}p{(\linewidth - 6\tabcolsep) * \real{0.3333}}@{}}

\toprule\noalign{}
\begin{minipage}[b]{\linewidth}\raggedright
Coefficient
\end{minipage} & \begin{minipage}[b]{\linewidth}\raggedleft
mle(), encrypted
\end{minipage} & \begin{minipage}[b]{\linewidth}\raggedleft
mle(), cleartext
\end{minipage} & \begin{minipage}[b]{\linewidth}\raggedright
Absolute Difference
\end{minipage} \\
\midrule\noalign{}
\endhead
\bottomrule\noalign{}

\tabularnewline
\caption{\label{tbl-cox-compare}Single-decrypter CKKS DLBCL Cox against
the same objective evaluated in the clear.}

\tabularnewline
\endlastfoot
GCB\_sig & -0.2638698 & -0.2638698 & \(4.76 \times 10^{-10}\) \\
LN\_sig & -0.2543587 & -0.2543587 & \(1.47 \times 10^{-10}\) \\
Prolif\_sig & 0.3031250 & 0.3031250 & \(2.26 \times 10^{-10}\) \\
BMP6 & 0.3036367 & 0.3036367 & \(7.38 \times 10^{-11}\) \\
MHC2\_sig & -0.3191459 & -0.3191459 & \(1.70 \times 10^{-10}\) \\

\end{longtable}

}

The encrypted fit agrees with the cleartext fit to within
\(4.76 \times 10^{-10}\) in every coefficient.

The example also shows that \texttt{mle()} itself needed no
modification: it issued its usual sequence of objective evaluations and
converged as it would have on cleartext data, unaware that each
evaluation was an encrypted round across three sites. Furthermore, the
patient records never left their sites and were never encrypted; each
site computed its partial log-likelihood on cleartext data it already
held, and encrypted that single scalar before returning it.

\subsection{Threshold variant: removing the single-decrypter
assumption}\label{sec-cox-threshold}

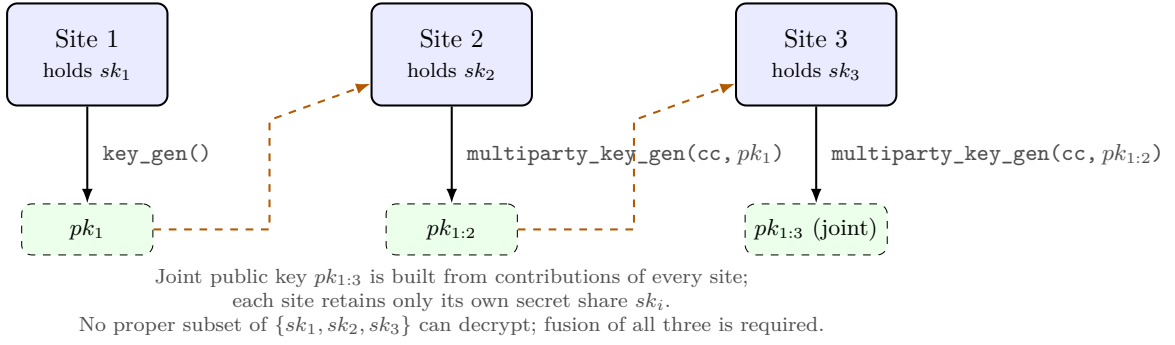
\begin{figure}[h]
  \centering
  \resizebox{\linewidth}{!}{\begin{tikzpicture}[
    node distance = 14mm and 18mm,
    site/.style   = {draw, rounded corners, thick, minimum width = 22mm,
                     minimum height = 14mm, align = center, fill = blue!8},
    pkbox/.style  = {draw, dashed, rounded corners, minimum width = 18mm,
                     minimum height = 7mm, align = center, fill = green!8,
                     font = \footnotesize},
    arr/.style    = {-{Latex[length=2.0mm]}, thick},
    arr2/.style   = {-{Latex[length=2.0mm]}, thick, color = orange!70!black, dashed},
    label/.style  = {font = \footnotesize, color = gray!60!black, align = center},
    every node/.append style = {font = \small}
  ]

  \node[site] (S1) at (0,    0) {Site 1\\\scriptsize{holds $sk_1$}};
  \node[site] (S2) at (5.0,  0) {Site 2\\\scriptsize{holds $sk_2$}};
  \node[site] (S3) at (10.0, 0) {Site 3\\\scriptsize{holds $sk_3$}};

  \node[pkbox] (PK1) at (0,   -2.4) {$pk_1$};
  \node[pkbox] (PK12) at (5.0, -2.4) {$pk_{1{:}2}$};
  \node[pkbox] (PK123) at (10.0, -2.4) {$pk_{1{:}3}$ (joint)};

  \draw[arr] (S1)  -- (PK1)
    node[midway, right=2pt, label] {\texttt{key\_gen()}};
  \draw[arr] (S2)  -- (PK12)
    node[midway, right=2pt, label] {\texttt{multiparty\_key\_gen(cc,\,$pk_1$)}};
  \draw[arr] (S3)  -- (PK123)
    node[midway, right=2pt, label] {\texttt{multiparty\_key\_gen(cc,\,$pk_{1{:}2}$)}};

  \draw[arr2] (PK1) -- ++(2.5, 0) -- ++(0.0, 1.5) -- (S2);
  \draw[arr2] (PK12) -- ++(2.5, 0) -- ++(0.0, 1.5) -- (S3);

  \node[align = center, font = \scriptsize, color = gray!60!black]
    at (5.0, -3.4)
    {Joint public key $pk_{1{:}3}$ is built from contributions of every site;\\
     each site retains only its own secret share $sk_i$.\\
     No proper subset of $\{sk_1, sk_2, sk_3\}$ can decrypt; fusion of all three is required.};
\end{tikzpicture}}
  \caption{Threshold key generation by chained
  \texttt{multiparty\_key\_gen()}. Each site contributes to the
  evolving joint public key while retaining only its own secret
  share. Encryption goes under the final joint public key
  $pk_{1\hbox{:}3}$; decryption requires every site to supply a
  partial decryption that the aggregator fuses. The
  \texttt{make\_threshold\_master()} constructor in
  \pkg{homomorpheR} runs this whole ceremony in one call.}
  \label{fig-tkg}
\end{figure}

The single-decrypter fit above leaves one structural assumption in
place: the master holds the secret key and could in principle decrypt
the per-site contributions before summing. The threshold variant removes
that assumption. The construction chains key generation across all
participating sites, each step run at the site by
\texttt{keygen\_round()}: site 1 calls \texttt{key\_gen()} and generates
an initial keypair, site 2 calls \texttt{multiparty\_key\_gen()} with
site 1's public key and produces \(pk_{1{:}2}\) and its own \(sk_2\),
site 3 calls it with \(pk_{1{:}2}\) and produces the joint
\(pk_{1{:}3}\) and its own \(sk_3\). The joint public key encrypts;
decryption requires every site to contribute a partial decryption that
the aggregator then fuses (the n-of-n decryption protocol). Figure
\ref{fig-tkg} illustrates the construction.

In code, the swap from single-decrypter to threshold is small: enable
\texttt{Feature\$MULTIPARTY} on the context, build fresh site objects,
and build the master with \texttt{make\_threshold\_master()} in place of
\texttt{make\_ckks\_master()}; the sites meet the context inside that
constructor, which runs the key-generation chain through them.

\begin{verbatim}
cc <- openfhe.R::fhe_context(scheme               = "CKKS",
                             multiplicative_depth = 1L,
                             scaling_mod_size     = 59L,
                             first_mod_size       = 60L,
                             batch_size           = 8L,
                             features = c(openfhe.R::Feature$MULTIPARTY))

workers_t <- Map(function(nm, d)
                     make_worker(name            = nm,
                                 data            = d,
                                 contribution_fn = local_cox_nll),
                 names(cox_data), cox_data)

master <- make_threshold_master(name           = "Aggregator",
                                crypto_context = cc,
                                sites          = workers_t)
\end{verbatim}

Two things differ from the single-decrypter setup. The sites are passed
\emph{to} the constructor rather than wired afterwards with
\texttt{set\_workers()}, because the joint public key is a function of
all of them and does not exist until the chain has run through each one.
These are also \emph{fresh} site objects, not the ones from the previous
section, which would be incompatible with this new protocol. The
\texttt{stats4::mle()} call is otherwise unchanged and is not repeated.

The threshold fit agrees with the same cleartext objective to within
\(1.73 \times 10^{-12}\) in every coefficient, yet no single remote
party can decrypt unilaterally: recovering each aggregate requires every
site's partial decryption. That is, the authority to decrypt is now
distributed.

\begin{longtable}[]{@{}
  >{\raggedright\arraybackslash}p{(\linewidth - 6\tabcolsep) * \real{0.1739}}
  >{\raggedleft\arraybackslash}p{(\linewidth - 6\tabcolsep) * \real{0.2464}}
  >{\raggedleft\arraybackslash}p{(\linewidth - 6\tabcolsep) * \real{0.2464}}
  >{\raggedright\arraybackslash}p{(\linewidth - 6\tabcolsep) * \real{0.3333}}@{}}
\toprule\noalign{}
\begin{minipage}[b]{\linewidth}\raggedright
Coefficient
\end{minipage} & \begin{minipage}[b]{\linewidth}\raggedleft
mle(), threshold
\end{minipage} & \begin{minipage}[b]{\linewidth}\raggedleft
mle(), cleartext
\end{minipage} & \begin{minipage}[b]{\linewidth}\raggedright
Absolute Difference
\end{minipage} \\
\midrule\noalign{}
\endfirsthead
\toprule\noalign{}
\begin{minipage}[b]{\linewidth}\raggedright
Coefficient
\end{minipage} & \begin{minipage}[b]{\linewidth}\raggedleft
mle(), threshold
\end{minipage} & \begin{minipage}[b]{\linewidth}\raggedleft
mle(), cleartext
\end{minipage} & \begin{minipage}[b]{\linewidth}\raggedright
Absolute Difference
\end{minipage} \\
\midrule\noalign{}
\endhead
\bottomrule\noalign{}
\tabularnewline
\caption{Threshold-CKKS DLBCL Cox against the same objective evaluated
in the clear.}\tabularnewline
\endlastfoot
GCB\_sig & -0.2638698 & -0.2638698 & \(1.73 \times 10^{-12}\) \\
LN\_sig & -0.2543587 & -0.2543587 & \(8.39 \times 10^{-13}\) \\
Prolif\_sig & 0.3031250 & 0.3031250 & \(5.27 \times 10^{-13}\) \\
BMP6 & 0.3036367 & 0.3036367 & \(1.37 \times 10^{-12}\) \\
MHC2\_sig & -0.3191459 & -0.3191459 & \(2.72 \times 10^{-13}\) \\
\end{longtable}

\section{Federated convex optimization with CVXR}\label{sec-cvxr}

The examples so far show that \emph{function-call} optimizers such as
\texttt{mle()} and \texttt{optim()} can compose with the threshold-FHE
channel when the objective decomposes as a sum over sites. The approach
does not generalize directly to arbitrary convex programs, e.g.~those
described in \pkg{CVXR} \citep{CVXRJSS}. The latter operates on a
\emph{symbolic} problem, not a callback: we cannot hand \pkg{CVXR} a
function that secretly does FHE on every call, because \pkg{CVXR} does
not \emph{call} our function: it canonicalizes the problem once and
ships the canonical form to a numerical solver.

When the global problem can be split into per-site local subproblems
each handled by \pkg{CVXR}, we can use the threshold-FHE channel only to
compute the cross-site \textbf{consensus update}. This is
\textbf{federated consensus ADMM}
\citep{BoydParikhChuPeleatoEckstein2011}. The local \pkg{CVXR} problem
can take advantage of \emph{disciplined parametric programming} (DPP)
\citep{AgrawalAmosBarrattBoyd2019} and stay efficient because the
\texttt{Parameter} values change but the problem structure does not.
During the consensus aggregation, the per-site \((x_k + u_k)\) vectors
are summed and averaged under encryption and the result is recovered via
threshold decryption.

\subsection{The global problem and its consensus
split}\label{the-global-problem-and-its-consensus-split}

We will illustrate the same \textbf{stratified Cox lasso} on the DLBCL
gene-expression cohort as the distributed Cox regression of
Section~\ref{sec-cox}. We used five published expression variables in
Section~\ref{sec-cox} but here we also include a larger number of
gene-expression predictors along with an L1 penalty. With \(N\) sites
(the molecular subgroups GCB, ABC, Type III), per-site standardized
design matrices \(X_k \in \mathbb{R}^{n_k
\times p}\), event times \(t_k\) and indicators \(\delta_k\), a shared
coefficient \(\beta \in \mathbb{R}^p\), and the per-stratum partial
log-likelihood \(\ell_k\) in Breslow form, the global problem is

\[
\min_{\beta \in \mathbb{R}^p}\;
-\sum_{k=1}^{N}\ell_k(\beta) \;+\; \lambda\,\lVert\beta\rVert_1.
\]

Because the partial likelihood factorizes additively across strata, the
consensus split \citep{BoydParikhChuPeleatoEckstein2011} is direct:
introduce per-site copies \(x_k \in \mathbb{R}^p\) and a single global
consensus \(z \in \mathbb{R}^p\),

\[
\min_{\{x_k\},\,z}\;
\sum_{k}\bigl(-\ell_k(x_k)\bigr)\,+\,\lambda\,\lVert z\rVert_1
\quad\text{s.t.}\quad x_k = z,\ \forall k.
\]

The augmented-Lagrangian iteration is

\[
\begin{aligned}
x_k^{t+1} &= \arg\min_{x}\, -\ell_k(x) + \tfrac{\rho}{2}\lVert x - z^t + u_k^t\rVert_2^2,\\
z^{t+1}   &= S_{\lambda/(N\rho)}\!\left(\tfrac{1}{N}\sum_k\bigl(x_k^{t+1} + u_k^t\bigr)\right),\\
u_k^{t+1} &= u_k^t + \bigl(x_k^{t+1} - z^{t+1}\bigr),
\end{aligned}
\]

where \(S_\tau(v) = \mathrm{sign}(v)\max(|v|-\tau, 0)\) is the
element-wise soft-thresholding operator (the proximal map of
\(\tau\lVert\cdot\rVert_1\)). The \(x\)-update is the per-site CVXR
solve; the \(z\)-update needs the cross-site average \(\bar w =
\tfrac{1}{N}\sum_k(x_k + u_k)\) which traverses the encrypted channel.
Soft-thresholding at the aggregator after decryption is closed-form and
adds no cryptographic depth.

The protocol is developed in two passes. The first runs the entire
federated fit \textbf{in the clear} to fix the target: standardize the
features, screen to a workable number of probes, and solve the consensus
ADMM with an ordinary unencrypted average, checking the result against
the centralized \pkg{CVXR} solve. The second puts it \textbf{under
threshold FHE}, replacing each cross-site sum --- the standardization
moments, the screening statistics, and the per-iteration consensus
average --- with one round of the encrypted-summation primitive, and
confirms that the encrypted fit reproduces the in-the-clear reference.

We display only relevant parts of the code here, read verbatim from the
pipeline installed with \pkg{homomorpheR}. The fit itself is precomputed
and the results shown here, so that we don't re-run the 112-iteration
encrypted ADMM. A reader can reproduce the fit end to end by rendering
the package vignette \texttt{cvxr-cox-lasso-dlbcl} with the environment
variable \texttt{HOMOMORPHER\_RECOMPUTE=true} set. Because threshold key
generation is randomized, however, computed coefficients will match the
shipped ones up to CKKS approximation noise.

\subsection{The federated fit in the
clear}\label{the-federated-fit-in-the-clear}

The three molecular subgroups are our sites. We load the cohort, split
it by subgroup, and keep each site's raw expression matrix, event times,
and status.

\begin{verbatim}
data(DLBCL,     package = "homomorpheR")  # 235 patients: survival + signatures
data(DLBCL_gex, package = "homomorpheR")  # 235 patients x 6416 Lymphochip probes
dlbcl <- DLBCL
dlbcl$Subgroup <- factor(dlbcl$Subgroup, levels = c("GCB","ABC","Type III"))
stopifnot(identical(as.character(dlbcl$ID), rownames(DLBCL_gex)))  # rows aligned

sites_raw <- lapply(levels(dlbcl$Subgroup), function(s) {
    idx <- which(dlbcl$Subgroup == s)
    list(name = s, X = DLBCL_gex[idx, ], time = dlbcl$time[idx],
         status = dlbcl$status[idx])
})
names(sites_raw) <- levels(dlbcl$Subgroup)
N_sites <- length(sites_raw)
N_total <- sum(vapply(sites_raw, function(s) nrow(s$X), 1L))
P_raw   <- ncol(DLBCL_gex)
\end{verbatim}

We standardize the features so the L1 penalty applies uniformly across
predictors on different scales. The pooled mean and variance are
additive over sites: site \(k\) contributes \(S_k = \sum_{i\in k} x_i\)
and \(Q_k = \sum_{i\in k} x_i^2\), from which
\(\mu = \tfrac{1}{N_{\mathrm{tot}}}\sum_k S_k\) and
\(\sigma^2 = \tfrac{1}{N_{\mathrm{tot}}}\sum_k Q_k - \mu^2\). In the
clear this is a pair of \texttt{colSums}.

\begin{verbatim}
pool_plain <- function(sites, n_total) {
    s <- Reduce(`+`, lapply(sites, function(s) colSums(s$X)))
    q <- Reduce(`+`, lapply(sites, function(s) colSums(s$X^2)))
    mu     <- s / n_total
    sigma2 <- pmax(q / n_total - mu^2, .Machine$double.eps)
    list(mu = mu, sigma = sqrt(sigma2))
}
## Each site standardizes its own rows with the pooled moments.
standardize <- function(sites, pool) lapply(sites, function(s) list(
    name = s$name,
    X    = sweep(sweep(s$X, 2, pool$mu, "-"), 2, pool$sigma, "/"),
    time = s$time, status = s$status))
pool      <- pool_plain(sites_raw, N_total)
sites_std <- standardize(sites_raw, pool)
\end{verbatim}

Rather than fit all \(p = 6416\) probes, we pre-screen to the
\(K = 100\) probes with the strongest univariate association with
survival. This keeps the memory footprint of the example manageable, and
it is the screen-then-fit pattern \citet{BayleFanLou2025} use on this
cohort. For probe \(g\) on stratum \(k\) in event-time order with risk
set \(R_i^{(k)}\) at event \(i\),
\(U_g^{(k)} = \sum_{i\in k,\,\delta_i=1}(X_{i,g} -
\overline{X}_{R_i^{(k)},g})\) is the score and
\(I_g^{(k)} = \sum_{i\in k,\,\delta_i=1}
\widehat{\mathrm{Var}}_{R_i^{(k)}}(X_{:,g})\) the information; both sum
across sites, and the screen ranks probes by
\(|Z_g| = |U_g|/\sqrt{I_g}\).

\begin{verbatim}
K <- 100L

score_info_at_zero <- function(X, time, status) {
    p <- ncol(X)
    ord <- order(time)
    X_o <- X[ord, , drop = FALSE]; stat_o <- status[ord]; time_o <- time[ord]
    n <- nrow(X_o); U <- numeric(p); I <- numeric(p)
    ## Risk set at event i: every row with time >= t_i, tied rows included.
    first <- match(time_o, time_o)
    for (i in seq_len(n)) {
        if (stat_o[i] == 1L) {
            risk <- X_o[first[i]:n, , drop = FALSE]
            mu_R <- colMeans(risk)
            U <- U + (X_o[i, ] - mu_R)
            I <- I + colSums(sweep(risk, 2, mu_R, "-")^2) / nrow(risk)
        }
    }
    list(U = U, I = I)
}

screen_plain <- function(sites, K) {
    UI <- lapply(sites, function(s) score_info_at_zero(s$X, s$time, s$status))
    U  <- Reduce(`+`, lapply(UI, `[[`, "U"))
    I  <- Reduce(`+`, lapply(UI, `[[`, "I"))
    Z  <- U / sqrt(pmax(I, .Machine$double.eps))
    order(abs(Z), decreasing = TRUE)[seq_len(K)]
}
## Each site keeps the screened probes of its own rows.
keep_probes <- function(sites, idx) lapply(sites, function(s) list(
    name = s$name, X = s$X[, idx],
    time = s$time, status = s$status))
top_idx  <- screen_plain(sites_std, K)
sites_KS <- keep_probes(sites_std, top_idx)
sigma_K  <- pool$sigma[top_idx]
\end{verbatim}

\texttt{sites\_KS} now holds each stratum on the \(K = 100\) screened
probes, and \texttt{sigma\_K} keeps their pooled SDs for the
back-transform to the original gene-expression scale.

The centralized stratified Cox-lasso fit --- our ground truth --- is a
single \pkg{CVXR} solve summing the per-stratum Breslow partial
likelihoods plus the L1 penalty.

\begin{verbatim}
LAMBDA <- 5
build_cox_breslow_nll <- function(beta_var, X_s, time_s, status_s) {
    ord <- order(time_s)
    X_o <- X_s[ord, , drop = FALSE]; stat_o <- status_s[ord]
    time_o <- time_s[ord]
    n <- nrow(X_o); eta_o <- X_o %*% beta_var
    ## Risk set at event i: every row with time >= t_i, tied rows included.
    first <- match(time_o, time_o)
    terms <- list()
    for (i in seq_len(n)) {
        if (stat_o[i] == 1L) {
            terms[[length(terms) + 1L]] <-
                log_sum_exp(eta_o[first[i]:n, 1]) - eta_o[i, 1]
        }
    }
    Reduce(`+`, terms)
}

beta_var <- Variable(K, name = "beta")
nll_per  <- lapply(sites_KS, function(s)
    build_cox_breslow_nll(beta_var, s$X, s$time, s$status))
agg_prob <- Problem(Minimize(Reduce(`+`, nll_per) +
                                 LAMBDA * p_norm(beta_var, 1)))
suppressMessages(suppressWarnings(
    psolve(agg_prob, solver = "CLARABEL", verbose = FALSE)))
agg_beta <- as.numeric(value(beta_var))
\end{verbatim}

The distributed fit splits this objective into per-site subproblems tied
by a consensus variable. Each local problem is built so DPP applies:
\(z\) and \(u\) enter as \texttt{Parameter}s whose values change every
ADMM iteration while the symbolic structure does not; \(X_k\), time,
status, and \(\rho\) are constants.

\begin{verbatim}
RHO <- 50

build_local <- function(X_k, time_k, status_k, rho) {
    p <- ncol(X_k)
    x  <- Variable(p); zp <- Parameter(p); up <- Parameter(p)
    nll <- build_cox_breslow_nll(x, X_k, time_k, status_k)
    aug <- (rho / 2) * sum_squares(x - zp + up)
    list(prob = Problem(Minimize(nll + aug)), x = x, zp = zp, up = up)
}
sites_problem <- lapply(sites_KS, function(s)
    build_local(s$X, s$time, s$status, RHO))
\end{verbatim}

The ADMM driver below is the whole federated algorithm, and it is
deliberately agnostic to \emph{how} the cross-site average is formed:
the \texttt{consensus} argument is a function of the per-site
\((x_k, u_k)\) vectors, and everything else --- the local \pkg{CVXR}
solves, the soft-threshold \(z\)-update, the dual update, the stopping
rule --- is ordinary \proglang{R}. We call it once now with an
unencrypted average and, unchanged, once more under encryption.

A note on the constants. We fix \(\rho = 50\) and a cap of 200
iterations, and stop when the absolute rule
\texttt{primal\ \textless{}\ TOL\ \&\&\ dual\ \textless{}\ TOL} holds,
with \texttt{TOL\ =\ 0.005}; the dual residual is
\(\rho\lVert z^{t+1}-z^t\rVert\).

\begin{verbatim}
MAX_ITER <- 200L; TOL <- 5e-3
soft_threshold <- function(v, tau) sign(v) * pmax(abs(v) - tau, 0)

run_admm <- function(sites_problem, consensus) {
    site_x <- replicate(N_sites, rep(0, K), simplify = FALSE)
    site_u <- replicate(N_sites, rep(0, K), simplify = FALSE)
    z_curr <- rep(0, K); trajectory <- list()
    for (iter in seq_len(MAX_ITER)) {
        for (i in seq_len(N_sites)) {
            value(sites_problem[[i]]$zp) <- z_curr
            value(sites_problem[[i]]$up) <- site_u[[i]]
            suppressMessages(suppressWarnings(
                psolve(sites_problem[[i]]$prob, solver = "CLARABEL",
                       verbose = FALSE)))
            site_x[[i]] <- as.numeric(value(sites_problem[[i]]$x))
        }
        w_avg  <- consensus(site_x, site_u)
        z_new  <- soft_threshold(w_avg, LAMBDA / (N_sites * RHO))
        site_u <- Map(function(u, x) u + (x - z_new), site_u, site_x)
        primal <- sqrt(mean(vapply(site_x, function(x) sum((x - z_new)^2), 0)))
        dual   <- RHO * sqrt(sum((z_new - z_curr)^2))
        z_curr <- z_new; trajectory[[iter]] <- z_new
        if (primal < TOL && dual < TOL) break
    }
    list(z = z_curr, trajectory = trajectory)
}

plain_consensus <- function(site_x, site_u)
    Reduce(`+`, Map(`+`, site_x, site_u)) / length(site_x)

ref        <- run_admm(sites_problem, plain_consensus)
z_ref      <- ref$z
n_iter_ref <- length(ref$trajectory)
\end{verbatim}

Here, in the clear, the consensus is just the average of the
\((x_k + u_k)\) vectors. The unencrypted ADMM converges in 112
iterations and matches the centralized \pkg{CVXR} fit to 0.0026 in
maximum absolute coefficient difference, so \texttt{agg\_beta} ---
equivalently \texttt{z\_ref} --- is the target the encrypted protocol
must reproduce.

\subsection{The same fit under threshold
FHE}\label{the-same-fit-under-threshold-fhe}

Apart from the key-generation round and the partial decryptions that
every fusion requires, four quantities cross a site boundary: the
standardization moments \((S_k, Q_k)\), the screening statistics
\((U^{(k)}, I^{(k)})\), and, at each ADMM iteration, the consensus sum
\(\sum_k (x_k + u_k)\) and the squared primal residuals
\(\lVert x_k - z\rVert^2\) of the stopping rule. \texttt{run\_admm}
forms the primal residual in the clear here; in a deployment it is one
more sum of the same kind. Each of the first three is a sum over sites,
so each becomes one round of the same threshold-FHE summation primitive
the master/worker fits used: every site encrypts its contribution under
the joint public key, the aggregator adds the encrypted contributions,
and the total is recovered by \(n\)-of-\(n\) partial decryption. For the
consensus, the encrypted total is first scaled by \(1/N\), so what is
decrypted is the average \(\bar w\). The local \pkg{CVXR} work and
\texttt{run\_admm} are untouched.

We use the same threshold infrastructure as
Section~\ref{sec-cox-threshold}: a CKKS context with
\texttt{Feature\$MULTIPARTY} and \texttt{make\_threshold\_master()},
which runs the key-generation chain through the sites. Each site
generates its own share and keeps it; only public keys move along the
chain, so no party --- the aggregator included --- ever holds the joint
secret.

So the sites have to exist before the aggregator does. Here we give each
subgroup a site object to hold its share.

\begin{verbatim}
cc <- fhe_context("CKKS",
                  multiplicative_depth = 1L,
                  scaling_mod_size     = 59L,
                  first_mod_size       = 60L,
                  batch_size           = 8192L,
                  features             = c(Feature$MULTIPARTY))

key_sites <- lapply(levels(dlbcl$Subgroup), function(s)
    make_worker(s, data = NULL, contribution_fn = function(data, theta) NULL))

master <- make_threshold_master("Aggregator",
                                crypto_context = cc, sites = key_sites)
\end{verbatim}

The standardization round is \texttt{pool\_plain} with the two
\texttt{colSums} encrypted: each site encrypts \(S_k\) and \(Q_k\), the
aggregator sums under encryption and threshold-decrypts the pooled
moments. The decrypted moments go back to the sites, which standardize
their own rows with them. (For brevity we treat \(N_{\mathrm{tot}}\) as
known to the aggregator; hiding the per-site head-counts is one more sum
of the same kind.)

\begin{verbatim}
## Site side. Each site forms its own moments and encrypts them under
## the joint key; what leaves is encrypted. Encrypting at the
## aggregator instead would mean handing it the per-site column sums in
## the clear first, which is the disclosure this round exists to avoid.
site_moments <- function(site, s)
    list(sum   = encrypt(site, colSums(s$X)),
         sumsq = encrypt(site, colSums(s$X^2)))

## Aggregator side. It receives only the encrypted moments, adds them,
## and decrypts the totals.
pool_encrypted <- function(master, parts, n_total, p_raw) {
    pooled_sum   <- decrypt(
        master, Reduce(`+`, lapply(parts, `[[`, "sum")),   len = p_raw)
    pooled_sumsq <- decrypt(
        master, Reduce(`+`, lapply(parts, `[[`, "sumsq")), len = p_raw)
    mu     <- pooled_sum / n_total
    sigma2 <- pmax(pooled_sumsq / n_total - mu^2, .Machine$double.eps)
    list(mu = mu, sigma = sqrt(sigma2))
}
moment_parts <- Map(site_moments, key_sites, sites_raw)   # at the sites
fhe_pool     <- pool_encrypted(master, moment_parts, N_total, P_raw)
pool_agree   <- list(mu    = max(abs(fhe_pool$mu    - pool$mu)),
                     sigma = max(abs(fhe_pool$sigma - pool$sigma)))

## The decrypted moments go back to the sites, which standardize their
## own rows with them.
sites_std_fhe <- standardize(sites_raw, fhe_pool)
\end{verbatim}

The encrypted moments agree with the unencrypted \texttt{pool} to
\ensuremath{4.6\times 10^{-16}} (mean) and
\ensuremath{3.2\times 10^{-15}} (SD). The screening round is
structurally identical --- the same \texttt{score\_info\_at\_zero}
summands \((U^{(k)}, I^{(k)})\), formed on the rows the sites have just
standardized, encrypted and summed the same way --- and it checks that
it selects the same probes, in the same order.

\begin{verbatim}
## Site side: compute the score and information at beta = 0 on the
## site's own rows, and encrypt both before returning them.
site_score_info <- function(site, s) {
    z <- score_info_at_zero(s$X, s$time, s$status)
    list(U = encrypt(site, z$U), I = encrypt(site, z$I))
}

## Aggregator side: it receives only the encrypted (U, I), adds them,
## and decrypts the totals.
screen_encrypted <- function(master, UI, p_raw, K) {
    U   <- decrypt(master, Reduce(`+`, lapply(UI, `[[`, "U")),
                   len = p_raw)
    I   <- decrypt(master, Reduce(`+`, lapply(UI, `[[`, "I")),
                   len = p_raw)
    Z   <- U / sqrt(pmax(I, .Machine$double.eps))
    order(abs(Z), decreasing = TRUE)[seq_len(K)]
}
UI_parts <- Map(site_score_info, key_sites, sites_std_fhe)   # at the sites
fhe_top  <- screen_encrypted(master, UI_parts, P_raw, K)
stopifnot(identical(fhe_top, top_idx))   # same probes, same order, as in the clear

## The decrypted screen goes back to the sites, which keep those probes.
sites_KS_fhe <- keep_probes(sites_std_fhe, fhe_top)
\end{verbatim}

The encrypted ADMM below runs on the design the two encrypted rounds
produced: each site's rows, standardized with the decrypted pooled
moments and restricted to the decrypted screen. The first movement's
design serves only as the reference. The consensus round is the only
piece left to encrypt. It mirrors \texttt{plain\_consensus}: encrypt
each \((x_k + u_k)\), sum the encrypted vectors, scale by \(1/N\) under
encryption (one multiplication by an unencrypted constant, one level of
multiplicative depth), and threshold-decrypt the length-\(K\) average.
Soft-thresholding stays in the clear at the aggregator.

\begin{verbatim}
## Site side: site k forms x_k + u_k and encrypts it. What leaves the
## site is encrypted.
site_consensus_term <- function(site, x_k, u_k)
    encrypt(site, x_k + u_k)

## Aggregator side: it receives only the encrypted terms, adds them,
## scales by 1/N, and decrypts the average.
aggregate_consensus <- function(master, cts, K)
    decrypt(master, Reduce(`+`, cts) * (1 / length(cts)), len = K)

## One consensus round: the site step at every site, then the
## aggregator step.
encrypted_consensus <- function(site_x, site_u)
    aggregate_consensus(master,
                        Map(site_consensus_term, key_sites, site_x, site_u),
                        K)

## The local problems, on the design the encrypted rounds produced.
sites_problem_fhe <- lapply(sites_KS_fhe, function(s)
    build_local(s$X, s$time, s$status, RHO))
fhe        <- run_admm(sites_problem_fhe, encrypted_consensus)
z_enc      <- fhe$z
trajectory <- fhe$trajectory
n_iter_enc <- length(trajectory)
\end{verbatim}

\texttt{run\_admm} is called unchanged, on local problems built from the
encrypted-round design and with \texttt{encrypted\_consensus} in place
of \texttt{plain\_consensus}. The encrypted setup, standardization, and
screening rounds described above are protocol work outside that driver.
The encrypted ADMM ran for 112 iterations and lands on the same
coefficients as the unencrypted run, differing by only
\ensuremath{2.4\times 10^{-7}} --- the CKKS approximation noise.
Table~\ref{tbl-cvxr-compare} compares it to the centralized fit on both
the standardized scale (where ADMM lives) and the back-transformed
original gene-expression scale.

{

\begin{longtable}[]{@{}lrr@{}}

\toprule\noalign{}
Scale & Max Absolute Difference & L1 Difference \\
\midrule\noalign{}
\endhead
\bottomrule\noalign{}

\tabularnewline
\caption{\label{tbl-cvxr-compare}Threshold-FHE consensus ADMM
vs.~centralized stratified Cox-lasso fit.}

\tabularnewline
\endlastfoot
standardized & 0.0026 & 0.0096 \\
original & 0.0048 & 0.0166 \\

\end{longtable}

}

The active set has 39 nonzero coefficients in the centralized fit and 39
in the encrypted ADMM fit; the intersection is 39 --- every probe
selected by the centralized fit is recovered by the encrypted
distributed protocol.

\begin{figure}

\centering{

\pandocbounded{\includegraphics[keepaspectratio]{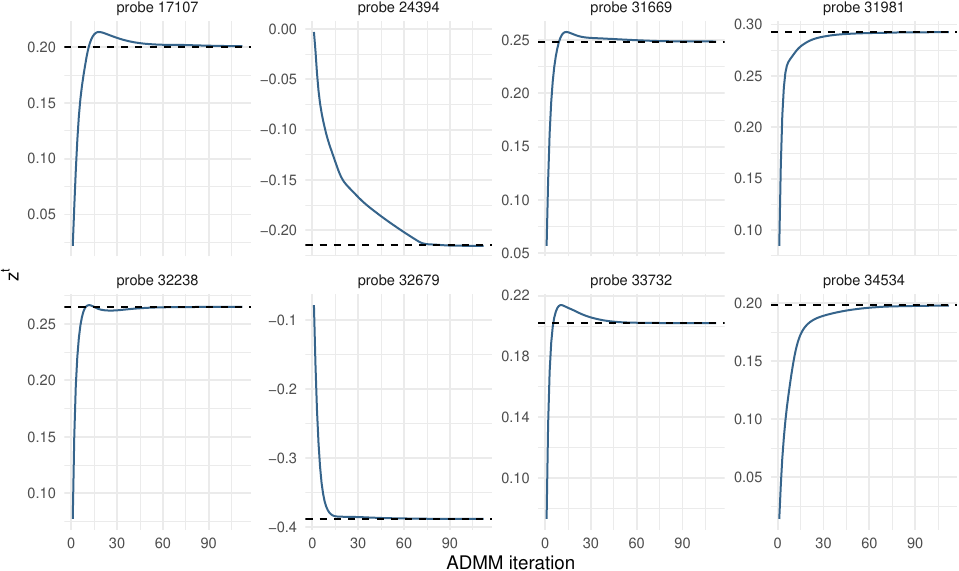}}

}

\caption{\label{fig-admm}Consensus trajectories for the eight
largest-magnitude coefficients in the encrypted stratified Cox lasso
fit. Solid lines are the encrypted ADMM iterates; horizontal dashed
lines are the centralized CVXR fit. Coefficients are on the standardized
scale.}

\end{figure}

Figure~\ref{fig-admm} shows the consensus trajectory \(z^t\) for the
eight probes with largest \(|z|\) at convergence, with the centralized
fit drawn as a horizontal reference. The encrypted iterates follow
ADMM's own trajectory toward the centralized CVXR solution; they are not
iterations of the centralized solver.

The aggregator observes the following, and nothing else: the pooled
per-probe sums \(\sum_k S_k\) and \(\sum_k Q_k\) (Round 1), the pooled
per-probe univariate Cox score \(U_g\) and information \(I_g\) (Round
2), the indices of the top-\(K\) screened probes (broadcast at the end
of Round 2), and, at every ADMM iteration, the cross-site average
\(\bar w\) before soft-thresholding and the primal residual. Each is an
aggregated population-level statistic over 235 patients, not
patient-level data; none of \(X_k\), the per-site \((S_k, Q_k)\), the
per-site \((U, I)\) contributions, or the per-site \((x_k + u_k)\)
vectors reaches the aggregator in cleartext. Decryption at every step is
\(n\)-of-\(n\) threshold; no single party --- including the aggregator
--- can recover any of these intermediate quantities unilaterally.

Two steps in this section would not exist in a real deployment. First,
the in-the-clear fit above and the centralized \pkg{CVXR} fit we compare
against both need all three sites' data in one place; we can run them
only because this is a demonstration. Second, a real deployment has to
choose \(\rho\) without that centralized fit. Choosing a constant such
as \(\rho\) is easy to overlook as a privacy risk because it looks like
setup rather than computation. It can be done without any real record:
for a simulated L2-regularized logistic regression, the
\texttt{cvxr-consensus-admm-dp} vignette picks \(\rho\) and the number
of iterations on a synthetic cohort built only from the number of sites,
their sizes, the covariate schema, and nominal effect sizes fixed in
advance.

\section{Additional protocol: federated similarity
retrieval}\label{sec-similarity}

The package vignette \texttt{similarity} applies the same threshold
primitives to a protocol whose shape differs from additive likelihood
aggregation. A public-space query embedding is broadcast in encrypted
form; each site computes encrypted similarity scores against its private
database; and the parties cooperate to decrypt the scores. We summarize
the construction because it exercises packed operations, rotations, and
threshold decryption, but do not treat it as a third validation study on
the scale of Section~\ref{sec-cox} or Section~\ref{sec-cvxr}.

Suppose site \(s\) has a private embedding model, database vectors
\(h_{sj}\) in its local space, and a compatibility map \(A_s\) that
relates its local space back to the public one, so that
\(A_s^\top h_{sj}\) is the public-space image of \(h_{sj}\). The map is
fitted from a public anchor cohort, embedded once with the public model
(the rows of \(H_0\)) and once with the site's private model (the rows
of \(H_s\)), by balancing anchor reconstruction against departure from
an isometry,

\[
\widehat A_s
= \arg\min_A
  \lVert H_s A - H_0 \rVert_F^2
  + \mu \lVert A^\top A-I\rVert_F^2.
\]

The endpoints range from an unconstrained least-squares map at
\(\mu=0\), computed in the package with a small ridge on the normal
equations, to the orthogonal Procrustes solution as \(\mu \to \infty\).
Both the fitted map and the local embeddings remain at the site. This
adapter is an application assumption, not a privacy mechanism: its
usefulness depends on the public anchors spanning the variation relevant
to future queries.

For a query \(q\), the raw score can be evaluated in either of two
algebraically equivalent forms,

\[
\langle A_s q, h_{sj}\rangle
= \langle q, A_s^\top h_{sj}\rangle.
\]

The first applies \(A_s\) to the encrypted query as a set of cleartext
diagonals, using rotations, cleartext multiplications, and additions for
the packed matrix-vector product. The second folds \(A_s^\top\) into the
database offline and reduces each online score to an encrypted dot
product. The latter has a shallower online computation but requires
recomputing the folded database whenever the adapter or database
changes. Once scores are cosines the two forms part company: the first
normalizes \(h_{sj}\) in the private space and the second normalizes
\(A_s^\top h_{sj}\) in the public space, and these agree only when
\(A_s\) is orthogonal. Encrypted division is not implicit in either
protocol.

A remote execution has four stages:

\begin{enumerate}
\def\labelenumi{\arabic{enumi}.}
\tightlist
\item
  The sites run joint key generation and retain their secret shares; the
  aggregator receives the joint public key. A second \(n\)-of-\(n\)
  ceremony among the sites produces the joint rotation keys that their
  scoring step uses.
\item
  The query owner encrypts \(q\) under the joint public key. The
  aggregator forwards only that encrypted query and public metadata.
\item
  Each remote site evaluates the first form against its local adapter
  and database, applying the adapter to the encrypted query and then
  taking one encrypted inner product per local embedding, and returns
  encrypted scores under the same joint key, with each candidate's local
  index and label in the clear. Neither the adapter, embeddings, nor
  cleartext scores are returned.
\item
  The sites supply partial decryptions, and the aggregator fuses them.
  Candidate indices and labels are protocol metadata and must be
  governed separately.
\end{enumerate}

The threat model is the same as in Section~\ref{sec-threat-model}. More
importantly, this protocol is \emph{not private top-\(k\) retrieval}.
The implementation threshold-decrypts every candidate score and ranks
the resulting numeric vector in the clear. The party receiving that
vector therefore learns all scores, their number, and whatever site or
record labels accompany them; repeated adaptive queries may reveal much
more. Hiding non-selected scores would require an encrypted comparison
or secure-selection protocol, which is not implemented here. Threshold
keys prevent unilateral decryption of intermediates, but they do not
change the disclosure selected by the final step.

\section{Two-party deployment and model extraction}\label{sec-threats}

The Cox and ADMM demonstrations of the previous sections fit into a
multi-site protocol where the aggregator coordinates many workers. We
now consider the two-party pattern: a single client querying a single
server, under the same honest-but-curious assumptions set out in
Section~\ref{sec-threat-model}. In the terms of that section this is a
single-key setting, not a threshold one, but the key is held by the
client, which owns the query, so the exposure runs the other way from
the single-decrypter bullet: what the key holder can recover is the
server's model, not another party's data. Both parties run in one
\proglang{R} session here, as everywhere in the paper.
\emph{Privacy-preserving inference} is the canonical example. A hospital
holds patient biomarkers and wants risk scores from a diagnostic lab
whose model is proprietary. The transport guarantee FHE provides is
straightforward: the hospital encrypts its query, the lab applies the
model homomorphically, the hospital decrypts the score, and the lab's
coefficients never leave its machine. The \texttt{secure-inference}
vignette implements that pipeline end to end and reports its agreement
with the cleartext computation; we do not repeat it here. Figure
\ref{fig-tpf} summarizes the data flow.

\begin{figure}[h]
  \centering
  \begin{tikzpicture}[
    node distance = 12mm and 24mm,
    box/.style    = {draw, rounded corners, thick, minimum width = 38mm,
                     minimum height = 24mm, align = left, font = \footnotesize},
    hospital/.style = {box, fill = blue!8},
    lab/.style    = {box, fill = orange!12},
    arr/.style    = {-{Latex[length=2.5mm]}, thick},
    arrlabel/.style = {font = \footnotesize, color = gray!60!black},
    every node/.append style = {font = \small}
  ]
  \node[hospital] (H) at (0, 0)
    {\textbf{Hospital (client)}\\
     \textit{cleartext:}\\
     \, biomarker $x = (x_1, \ldots, x_p)$\\
     \, decrypted score\\
     \textit{key material:}\\
     \, public key $pk$\\
     \, secret key $sk$};

  \node[lab] (L) at (8.4, 0)
    {\textbf{Lab (server)}\\
     \textit{cleartext:}\\
     \, model coefficients $w, b$\\
     \textit{key material:}\\
     \, public key $pk$ only\\
     \textit{view:}\\
     \, ciphertexts in / out};

  \draw[arr] ($(H.east) + (0, 0.3)$) -- ($(L.west) + (0, 0.3)$)
    node[midway, above, arrlabel] {$E_{pk}(x)$ \scriptsize(encrypted query)};
  \draw[arr] ($(L.west) + (0, -0.4)$) -- ($(H.east) + (0, -0.4)$)
    node[midway, below, arrlabel] {$E_{pk}(\text{score}) = w\cdot E_{pk}(x) + b$};

  \node[align = center, font = \scriptsize, color = gray!60!black]
    at (4.2, -2.4)
    {\textbf{Transport guarantees.}
    Hospital never sends $x$ in cleartext;
    lab never sends $w, b$ to hospital.\\
    \textbf{Residual exposure.} Hospital can still recover $w, b$
    by adaptive querying (see Section~\ref{sec-threats}).};
\end{tikzpicture}
  \caption{Two-party deployment under FHE. The hospital encrypts
  its biomarker vector under its public key, sends the
  encrypted vector to the lab, the lab applies its proprietary
  weighted-sum scoring to it while it stays encrypted, and returns an
  encrypted score that the hospital decrypts. The biomarker
  values never appear in cleartext on the lab's machine, and the
  lab's coefficients are never sent to the hospital. While these are
  necessary for any model-as-a-service deployment, they are
  not sufficient conditions, and can be attacked to reveal information.}
  \label{fig-tpf}
\end{figure}

The simplest case is a linear model with four biomarkers and a bias; the
hospital recovers every coefficient with five basis-vector queries:

\begin{align*}
\mathbf{e}_0 &= (0,0,0,0) \;\Rightarrow\; \text{score} = b, \\
\mathbf{e}_j &= \text{$j$-th unit vector} \;\Rightarrow\; \text{score} = w_j + b
\quad (j = 1, \ldots, 4).
\end{align*}

When run with the scoring function, the five probes --- one bias probe
plus one per biomarker --- the attack recover \(b\) and \(w\) to within
\ensuremath{1.2\times 10^{-13}}, which is CKKS precision. FHE does not
impede this extraction.

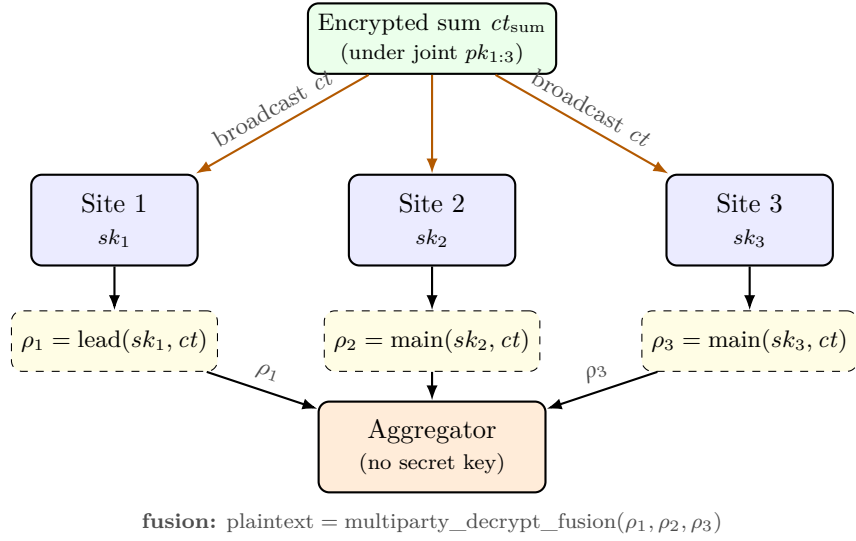
\begin{figure}[h]
  \centering
  \begin{tikzpicture}[
    node distance = 12mm and 18mm,
    site/.style    = {draw, rounded corners, thick, minimum width = 22mm,
                      minimum height = 12mm, align = center, fill = blue!8},
    aggr/.style    = {draw, rounded corners, thick, minimum width = 30mm,
                      minimum height = 12mm, align = center, fill = orange!15},
    cipher/.style  = {draw, rounded corners, thick, minimum width = 26mm,
                      minimum height = 8mm, align = center, fill = green!8,
                      font = \footnotesize},
    partial/.style = {draw, dashed, rounded corners, minimum width = 22mm,
                      minimum height = 8mm, align = center, fill = yellow!12,
                      font = \footnotesize},
    arr/.style     = {-{Latex[length=2.0mm]}, thick},
    arrL/.style    = {-{Latex[length=2.0mm]}, thick, color = orange!70!black},
    label/.style   = {font = \footnotesize, color = gray!60!black, align = center},
    every node/.append style = {font = \small}
  ]
  \node[cipher] (CT) at (0, 0)
    {Encrypted sum $\mathit{ct}_{\text{sum}}$\\\scriptsize (under joint $pk_{1{:}3}$)};

  \node[site] (S1) at (-4.2, -2.4) {Site 1\\\scriptsize{$sk_1$}};
  \node[site] (S2) at (0,    -2.4) {Site 2\\\scriptsize{$sk_2$}};
  \node[site] (S3) at ( 4.2, -2.4) {Site 3\\\scriptsize{$sk_3$}};

  \node[aggr] (AG) at (0, -5.4)
    {Aggregator\\\scriptsize (no secret key)};

  \node[partial] (P1) at (-4.2, -4.0) {$\rho_1 = \text{lead}(sk_1, \mathit{ct})$};
  \node[partial] (P2) at (0,    -4.0) {$\rho_2 = \text{main}(sk_2, \mathit{ct})$};
  \node[partial] (P3) at ( 4.2, -4.0) {$\rho_3 = \text{main}(sk_3, \mathit{ct})$};

  \draw[arrL] (CT) -- (S1) node[midway, above, sloped, label] {broadcast $\mathit{ct}$};
  \draw[arrL] (CT) -- (S2);
  \draw[arrL] (CT) -- (S3) node[midway, above, sloped, label] {broadcast $\mathit{ct}$};

  \draw[arr] (S1) -- (P1);
  \draw[arr] (S2) -- (P2);
  \draw[arr] (S3) -- (P3);

  \draw[arr] (P1) -- (AG) node[midway, sloped, above, label] {$\rho_1$};
  \draw[arr] (P2) -- (AG);
  \draw[arr] (P3) -- (AG) node[midway, sloped, above, label] {$\rho_3$};

  \node[align = center, font = \scriptsize, color = gray!60!black]
    at (0, -6.4)
    {\textbf{fusion:} $\text{plaintext} = \mathrm{multiparty\_decrypt\_fusion}(\rho_1, \rho_2, \rho_3)$};
\end{tikzpicture}
  \caption{Threshold partial-decryption fusion. The aggregator
  broadcasts the encrypted sum to every site; each site $i$
  computes a partial decryption using its own secret share
  $sk_i$ (\texttt{multiparty\_decrypt\_lead} for the first site,
  \texttt{multiparty\_decrypt\_main} for the rest); the aggregator
  fuses the partials via \texttt{multiparty\_decrypt\_fusion} to
  recover the sum in the clear. n-of-n: any subset of the partials
  is insufficient.}
  \label{fig-tdf}
\end{figure}

\subsection{Layered defenses}\label{layered-defenses}

A deployment might consider layering defenses \emph{on top of} the FHE
transport for a scoring function to stay proprietary. The four that
compose with the \pkg{openfhe.R} primitives are:

\begin{itemize}
\tightlist
\item
  \textbf{Output perturbation.} Adding calibrated noise to the decrypted
  score --- the differential-privacy mechanism --- turns each query into
  a noisy observation rather than an exact readout. Extraction accuracy
  degrades with the noise variance, and a query budget can be accounted
  for explicitly. Section~\ref{sec-dp} and the vignettes
  \texttt{cox-threshold-dp} and \texttt{cvxr-consensus-admm-dp} explore
  a related composition in a multi-site fit: each site adds an
  independent Gaussian draw to its contribution before encryption, so
  that the decrypted aggregate carries noise of a specified variance.
\item
  \textbf{Query-budget accounting.} Capping the number of queries a
  client can issue in a session bounds how much of the model can leak
  before the budget is exhausted. Composes with output perturbation
  rather than replacing it.
\item
  \textbf{Threshold key generation.} Splitting the secret key across
  several parties via \texttt{make\_threshold\_master()} (or a
  client-server analogue) changes \emph{who} can decrypt and under what
  audit trail: the hospital alone can no longer open a score, since
  every release needs the cooperation of every key-holder. This merely
  controls release and does not prevent extraction if the key-holders
  routinely approve each query as it arrives. Figure \ref{fig-tdf} shows
  the partial-decryption fusion protocol.
\item
  \textbf{Non-linear scoring.} A linear model is the trivial extraction
  case. CKKS supports polynomial arithmetic, and Chebyshev
  approximations evaluate transcendental functions under encryption:
  \texttt{eval\_chebyshev\_function} for a general function and
  \texttt{eval\_logistic} for the logistic sigmoid. The
  \texttt{encrypted-regression} vignette of \pkg{homomorpheR}
  demonstrates a logistic-regression prediction pipeline whose scoring
  function \(\sigma(\beta^\top x)\) is computed entirely in the
  encrypted domain --- coefficients applied to encrypted features,
  sigmoid evaluated as a degree-16 Chebyshev polynomial, decryption only
  at the very end. This just raises the cost of extraction rather than
  preventing it.
\end{itemize}

\subsection{Security limitations at a
glance}\label{security-limitations-at-a-glance}

The package primitives address where cleartext appears and who can
decrypt; they do not by themselves make a protocol private, available,
or robust to malicious behavior. Table~\ref{tbl-toolbox} collects the
main limitations used in interpreting the examples.

\begin{longtable}[]{@{}
  >{\raggedright\arraybackslash}p{(\linewidth - 4\tabcolsep) * \real{0.3333}}
  >{\raggedright\arraybackslash}p{(\linewidth - 4\tabcolsep) * \real{0.3333}}
  >{\raggedright\arraybackslash}p{(\linewidth - 4\tabcolsep) * \real{0.3333}}@{}}
\toprule\noalign{}
\begin{minipage}[b]{\linewidth}\raggedright
Limitation
\end{minipage} & \begin{minipage}[b]{\linewidth}\raggedright
Package stance
\end{minipage} & \begin{minipage}[b]{\linewidth}\raggedright
Operational consequence
\end{minipage} \\
\midrule\noalign{}
\endfirsthead
\toprule\noalign{}
\begin{minipage}[b]{\linewidth}\raggedright
Limitation
\end{minipage} & \begin{minipage}[b]{\linewidth}\raggedright
Package stance
\end{minipage} & \begin{minipage}[b]{\linewidth}\raggedright
Operational consequence
\end{minipage} \\
\midrule\noalign{}
\endhead
\bottomrule\noalign{}
\tabularnewline
\caption{Principal limitations of the demonstrations and their
operational consequences. These qualifications apply in addition to the
local-versus-remote distinction in
Section~\ref{sec-threat-model}.}\label{tbl-toolbox}\tabularnewline
\endlastfoot
A single secret key permits unilateral decryption & Use \(n\)-of-\(n\)
key generation and partial-decryption fusion where that trust assumption
is unacceptable & Every decryption requires every enrolled party to be
online \\
CKKS has finite depth and approximate arithmetic & Design shallow
computations and select parameters for the declared workload;
bootstrapping is exposed but not used in the main examples & Precision
and level consumption must be checked for each computation \\
Decrypted outputs can leak data or model parameters & Treat release
policy, query authorization, and accounting as separate protocol
requirements & Threshold keys govern who approves a release; they do not
reduce what an approved output reveals \\
Site-level noise does not automatically provide differential privacy &
Require an explicit neighboring relation, sensitivity bound, clipping
rule, and accountant & The exploratory mechanism in Section~\ref{sec-dp}
is not a DP guarantee \\
Honest-but-curious sites may return incorrect values & No malicious-site
protection is implemented & Verifiable computation, commitments, or
zero-knowledge proofs would be additional protocol layers \\
\end{longtable}

\subsection{Cryptographic security
note}\label{cryptographic-security-note}

The cryptographic security of CKKS --- the approximate scheme this paper
uses for all real-valued computation --- has been the subject of active
research. \citet{LiMicciancio2021} showed that a party who observes both
the encrypted and decrypted outputs of a computation can recover the
secret key from the noise pattern, and proposed a mitigation
(IND-CPA\(^{\mathrm{D}}\)) based on adding Gaussian noise to the
decrypted output before releasing it. OpenFHE implements this
mitigation. Subsequent work
\citep{GuoNabokovSuvantoJohansson2024, CheonChoePasselegueStehle2024, ChecriSirdeyBoudguigaBultel2024}
showed that the mitigation can be circumvented when the parameters are
chosen for one computation but a different computation is actually
performed; \citet{AlexandruAlBadawiMicciancioPolyakov2024} propose an
application-aware framework that validates at evaluation time that the
computation matches the declared parameter profile. As of this writing,
the vendored OpenFHE v1.5.1 release does not include that framework.

The demonstrations here therefore use OpenFHE's default CKKS mode, under
the honest-but-curious model of Section~\ref{sec-threat-model}, and
claim no IND-CPA\(^{\mathrm{D}}\) security. In particular we do not set
\texttt{decryption\_noise\_mode\ =\ DecryptionNoiseMode\$NOISE\_FLOODING\_DECRYPT},
run the \texttt{ExecutionMode\$EXEC\_NOISE\_ESTIMATION} pass, or
generate a fresh key pair for an evaluation context, which are the steps
OpenFHE's two-phase experimental mode requires. A deployment that
releases CKKS decryptions to a party that also observes the
corresponding encrypted values should follow OpenFHE's noise-flooding
guidance for its exact computation and query budget. The
\texttt{openfhe.R} binding exposes the relevant parameters.

\section{Exploratory composition with Gaussian noise}\label{sec-dp}

The package vignettes \texttt{cox-threshold-dp} and
\texttt{cvxr-consensus-admm-dp} explore a mechanical composition: each
site adds an independent Gaussian draw before encryption so that the
aggregated perturbation has a specified variance, and a zCDP accountant
\citep{BunSteinke2016} composes the nominal per-query cost. Note that
these do not provide a differential-privacy guarantee; that would
require a neighboring-dataset definition, clipping or boundedness, a
proved sensitivity \(\Delta\), and an accountant covering every release
and data-dependent tuning decision \citep{DworkRoth2014}. The vignettes
set \(\Delta=1\) only as a placeholder for studying the software path.
Their \(\varepsilon\) calculations therefore must not be interpreted as
privacy claims.

What they show is that the two mechanisms compose: the site-side draws
are encrypted with the contributions and only their sum is decrypted, so
the decrypted aggregate carries noise of the specified variance; and at
zero variance the fit agrees with the cleartext stratified Cox fit in
the one case and with the centralized \pkg{CVXR} fit in the other. The
statistical interaction is less favorable. In the Cox example, BFGS
computes its gradient by finite differences of the noisy objective;
Nelder--Mead uses no gradient. Noise small enough for the optimizer to
tolerate still yields large nominal privacy losses under the placeholder
sensitivity, while larger noise materially destabilizes the fit.

\section{Related work}\label{sec-related}

This demonstrations presented here involve cryptographic computation in
R, so we begin by sketching the most related work in R.

\subsection{Adjacent work in R}\label{adjacent-work-in-r}

The most direct predecessor for this work is \pkg{distcomp}
\citep{NarasimhanRubinGrossBenderskyLavori2017, distcomp}, which the
present author developed in 2017. \pkg{distcomp} ships a distributed
stratified Cox demonstration with the same additive likelihood
decomposition this paper uses, on top of the \pkg{opencpu} REST
infrastructure \citep{Ooms2014}. The site summaries in \pkg{distcomp}
travel in cleartext between sites and the master; the present work
replaces that transport with the threshold-FHE channel. Mathematically
the two systems run the same algorithm; the cryptographic layer is the
only difference.

Earlier versions of \pkg{homomorpheR} \citep{homomorpheR} (first
released 2015) supported only additively homomorphic encryption, and
shipped a distributed stratified-Cox demonstration on that basis --- the
direct statistical predecessor of the multi-party Cox demonstrations
revisited under CKKS threshold FHE in Section~\ref{sec-cox}. The
\pkg{homomorpheR} 1.0 release described here builds on the
\pkg{openfhe.R}-backed class hierarchy instead; the Paillier primitives
are frozen and kept only for existing users.

DataSHIELD is a long-running \proglang{R}-based federated-analysis
platform for biomedical research, started in 2009
\citep{WolfsonDataSHIELD2010, MurtaghDataSHIELD2012, WallaceDataSHIELD2014, GayeDataSHIELD2014}
and synthesized in the 2025 treatment of
\citet{AvraamDataSHIELDDisclosure2025}. \pkg{DataSHIELD} is an umbrella
project covering more than twenty community packages. Its analysis
controls mitigate disclosure risk via output filtering and
aggregate-only access rather than encryption;
\citet{AvraamDataSHIELDDisclosure2025} name homomorphic encryption as a
future direction in which the DataSHIELD analysis controls are expected
to develop. The distributed-survival package \texttt{dsSurvival} and the
synthesis-and-aggregation work in \citet{AvraamDataSHIELDDisclosure2025}
are the closest points of comparison: they use related master/worker
decompositions, while the present work focuses on protecting
intermediate aggregates cryptographically.

Two \proglang{R} packages named \pkg{HomomorphicEncryption} also exist:
one on CRAN \citep{HomomorphicEncryptionR}, a pure \proglang{R}
implementation aimed at introducing the schemes rather than at
computation on real data, and one described in
\citet{AslettEsperancaHolmes2015}, which has a more performant C++
implementation using \proglang{OpenMP} threads but for selected
platforms. Neither exposes the combination of CKKS slot packing,
bootstrapping, and multiparty key generation used in this paper.

The Python-side sibling of \pkg{openfhe.R} is \pkg{openfhe-python}
\citep{OpenFHEPython}, also developed against the OpenFHE C++ library. A
higher-level Python counterpart is under development, but all
implementations and results reported here are the \proglang{R} versions.

\subsection{Cryptographic encrypted distributed
optimization}\label{cryptographic-encrypted-distributed-optimization}

Several groups have investigated combining homomorphic encryption with
distributed convex optimization since the late 2010s. The constructions
differ in cryptographic primitive, architecture, and threat model.

\begin{itemize}
\tightlist
\item
  \citet{LuZhu2018} layer Paillier on top of the gradient transmissions
  of an existing distributed optimization algorithm. The construction is
  fully decentralized and the privacy guarantee is that no agent learns
  another agent's private parameters from the encrypted values it
  observes. Limited by Paillier's additive-only homomorphism.
\item
  \citet{AlexandruGatsisPappas2017} give a Paillier-based two-party
  cloud quadratic-program protocol --- client and untrusted cloud
  worker, with a Damgård--Geisler--Krøigaard comparison sub-protocol for
  the active-set update. The architecture is single-cloud rather than
  peer-to-peer.
\item
  \citet{AlexandruTsiamisPappas2021} extend the same line to a
  CKKS-encrypted Lasso problem in which \(M\) data owners encrypt their
  data and ship the encrypted values to a cloud server that solves the
  regression problem under encryption. The soft-thresholding step is
  approximated by a Chebyshev polynomial, conceptually similar to the
  Chebyshev-sigmoid pattern in the \texttt{encrypted-regression}
  vignette of \pkg{homomorpheR}.
\item
  \citet{BinfetAdamekSchluterSchulzeDarup2023} take the consensus-ADMM
  template of \citet{BoydParikhChuPeleatoEckstein2011} and instantiate
  it under leveled CKKS with multiple cryptosystem instances and key
  switching between them. Their case study is privacy-preserving
  cooperative control of a multi-agent robot formation. The
  consensus-ADMM demonstration in Section~\ref{sec-cvxr} is in the same
  architectural family but uses n-of-n threshold key generation rather
  than multi-instance key switching, generic \pkg{CVXR} symbolic
  problems rather than a hand-coded quadratic update, and OpenFHE's
  default \texttt{HEStd\_128\_classic} security profile.
\end{itemize}

Related work in the malicious setting includes \textbf{Helen}
\citep{ZhengPopaGonzalezStoica2019}, which combines SPDZ-style MPC with
zero-knowledge proofs to run ADMM with malicious-site protection. It
establishes what is achievable in the malicious setting, at the cost of
the zero-knowledge proof machinery that setting requires; the present
work stays in the honest-but-curious setting and provides no
malicious-site protection. The differential-privacy companion on the
consensus-ADMM side is \citet{CyffersEvenBelletMassoulie2022}.

\subsection{Encrypted retrieval and representation
alignment}\label{encrypted-retrieval-and-representation-alignment}

The retrieval protocol of Section~\ref{sec-similarity} has near
relatives that combine some but not all of its ingredients.
\citet{ZhaoFRAG2024} (FRAG) let mutually distrusting parties run
approximate nearest-neighbor search over encrypted vectors under a
single-key homomorphic encryption protocol. \citet{MaoFedE4RAG2025}
(FedE4RAG) train client-side retrieval-augmented-generation retrievers
by federated learning, with knowledge distillation between the server
and client models and homomorphic encryption of the model parameters.
\citet{YaoPPER2024} build privacy-preserving entity resolution across
healthcare databases on AMPPERE, a secure computation model that uses
homomorphic encryption.

The worked protocol in Section~\ref{sec-similarity} combines
site-private models, a per-site compatibility adapter fitted on public
anchors, threshold CKKS, and two algebraically equivalent deployment
forms. It is presented as an integration example rather than a
systematic novelty claim; the package vignette supplies its synthetic
software checks.

\section{Discussion}\label{sec-discussion}

The two packages address a specific technical problem: expose OpenFHE's
primitives to R and provide master/worker classes through which
statistical solvers can use an encrypted aggregation channel.
Section~\ref{sec-cox} and Section~\ref{sec-cvxr} are the substantial
instances --- \texttt{stats4::mle()} driving a distributed Cox fit, and
a \pkg{CVXR} Cox-lasso by consensus ADMM --- and
Section~\ref{sec-similarity} and Section~\ref{sec-threats} add a
retrieval protocol and a two-party prediction pipeline of different
shapes, with more limited validation.

Two further demonstrations are developed only in the package vignettes.
The \texttt{privacy-preserving-aggregation} vignette shows exact integer
counting under single-key BFV, written directly against \pkg{openfhe.R},
and \texttt{query-count-threshold} applies the same counting criterion
to different simulated sites under threshold BFV, through the
master/worker classes used above. Because BFV is exact, the recovered
counts are integers rather than approximations.

Production use requires a federated-analysis platform. \pkg{DataSHIELD}
is one; the stack of the Observational Health Data Sciences and
Informatics consortium (\textbf{OHDSI}, \citet{HripcsakOHDSI2015}) is
another. Each supplies the governance, audit, communication
infrastructure, and identity management a deployment needs, and the
packages here could supply an encrypted aggregation layer within them.
The most natural next step is integration with one of those platforms
--- most plausibly \pkg{DataSHIELD} or OHDSI, whose analysis-function
signatures and master/worker patterns match the ones used here.

The \pkg{homomorpheR} vignettes carry fuller versions of the worked
protocols.

\subsection{Limitations}\label{limitations}

We make no claims on performance: the examples use modest cohort sizes
and parameters chosen for reproducible runs, and the paper does not
isolate R-binding overhead, network cost, or scaling with parties and
slots. Addressing real communication is also out of scope.
\pkg{openfhe.R} provides \texttt{fhe\_serialize()} and
\texttt{fhe\_deserialize()}, but authentication, replay protection,
identity, auditing, and endpoint key storage belong to the deployment.
Finally, the honest-but-curious model does not detect incorrect
encrypted contributions or malformed partial decryptions. Commitments,
verifiable computation, or zero-knowledge proofs would be additional
layers, not features of the present package.

\section{Computational details}\label{sec-comp}

The results in this paper were obtained using \proglang{R} 4.6.1 on
macOS Tahoe 26.6.2 (aarch64-apple-darwin23), with the \pkg{openfhe.R}
package 1.5.1.2 linked against OpenFHE v1.5.1, and \pkg{homomorpheR}
version 1.0. Table~\ref{tbl-sessioninfo} lists the versions of every
package the examples load.

{

\begin{longtable}[]{@{}ll@{}}

\toprule\noalign{}
Package & Version \\
\midrule\noalign{}
\endhead
\bottomrule\noalign{}

\tabularnewline
\caption{\label{tbl-sessioninfo}Package versions used to produce the
results in this paper.}

\tabularnewline
\endlastfoot
openfhe.R & 1.5.1.2 \\
homomorpheR & 1.0 \\
survival & 3.8.12 \\
CVXR & 1.9.2 \\
ggplot2 & 4.0.3 \\
S7 & 0.2.2 \\
knitr & 1.52 \\

\end{longtable}

}

\proglang{R}, \pkg{openfhe.R} and \pkg{homomorpheR} are available from
the \href{https://CRAN.R-project.org/}{Comprehensive R Archive Network
(CRAN)}.

A \proglang{Python} counterpart to \pkg{homomorpheR} called
\pkg{homomorphepy} bundling the same examples is available on
\href{https://github.com/bnaras/homomorphepy}{GitHub}.

\section{Use of AI}\label{use-of-ai}

AI models were used in developing both packages, doing literature
search, formatting the manuscript, all under the author's direction. The
author specified the design, reviewed the generated output, and is
responsible for all content. All results reported are from actual
computation.

\section{Acknowledgements}\label{acknowledgements}

This work was supported by Weill Cancer Hub West.

\bibliography{bibliography.bib}

\end{document}